\documentclass[a4paper,11pt]{article}
\pdfoutput=1 

\usepackage{amssymb,amsmath}
\usepackage{graphicx}
\usepackage{color}
\usepackage{float}
\usepackage{empheq}
\usepackage[font=footnotesize]{caption}
\usepackage[font=footnotesize]{subcaption}
\usepackage{hyperref}
\usepackage{appendix}
\usepackage{fullpage}

\newcommand{\dr}{\mathrm{d}r}
\newcommand{\dd}{\mathrm{d}}
\newcommand{\ep}{\epsilon}
\newcommand*\widefbox[1]{\fbox{\hspace{2em}#1\hspace{2em}}}

\DeclareMathOperator{\arctanh}{arctanh}
\DeclareMathOperator{\nlog}{ln}
\newcommand{\be}{\begin{equation}}
\newcommand{\bea}{\begin{eqnarray}}
\newcommand{\eea}{\end{eqnarray}}
\newcommand{\ee}{\end{equation}}

\begin{document}
\begin{titlepage}
\bigskip

\bigskip\bigskip\bigskip\bigskip

\centerline{\Large \bf {Geometry of the Infalling Causal Patch}}

\bigskip\bigskip\bigskip\bigskip

\centerline{\bf Ben Freivogel, Robert A. Jefferson, Laurens Kabir, and I-Sheng Yang}
\medskip
\centerline{\small ITFA and GRAPPA, Universiteit van Amsterdam}
\centerline{\small Science Park 904, 1098 XH, Amsterdam, the Netherlands}

\bigskip\bigskip\bigskip\bigskip

\begin{abstract} 
The firewall paradox states that an observer falling into an old black hole must see a violation of unitarity, locality, or the equivalence principle. Motivated by this remarkable conflict, we analyze the causal structure of black hole spacetimes in order to determine whether all the necessary ingredients for the paradox fit within a single observer's causal patch. We particularly focus on the question of whether the interior partner modes of the outgoing Hawking quanta can, in principle, be measured by an infalling observer. Since the relevant modes are spread over the entire sphere, we answer a simple geometrical question: can any observer see an entire sphere behind the horizon?

We find that for all static black holes in 3+1 and higher dimensions, with any value of the cosmological constant, no single observer can see both the early Hawking radiation and the interior modes with low angular momentum. We present a detailed description of the causal patch geometry of the Schwarzschild black hole in 3+1 dimensions, where an infalling observer comes closest to being able to measure the relevant modes.

\end{abstract}

\end{titlepage}

\noindent\rule{\textwidth}{0.5pt}
\tableofcontents
\noindent\rule{\textwidth}{0.5pt}

\section{Introduction}
\begin{figure}[!h]
\centering
\includegraphics[width=0.49\textwidth]{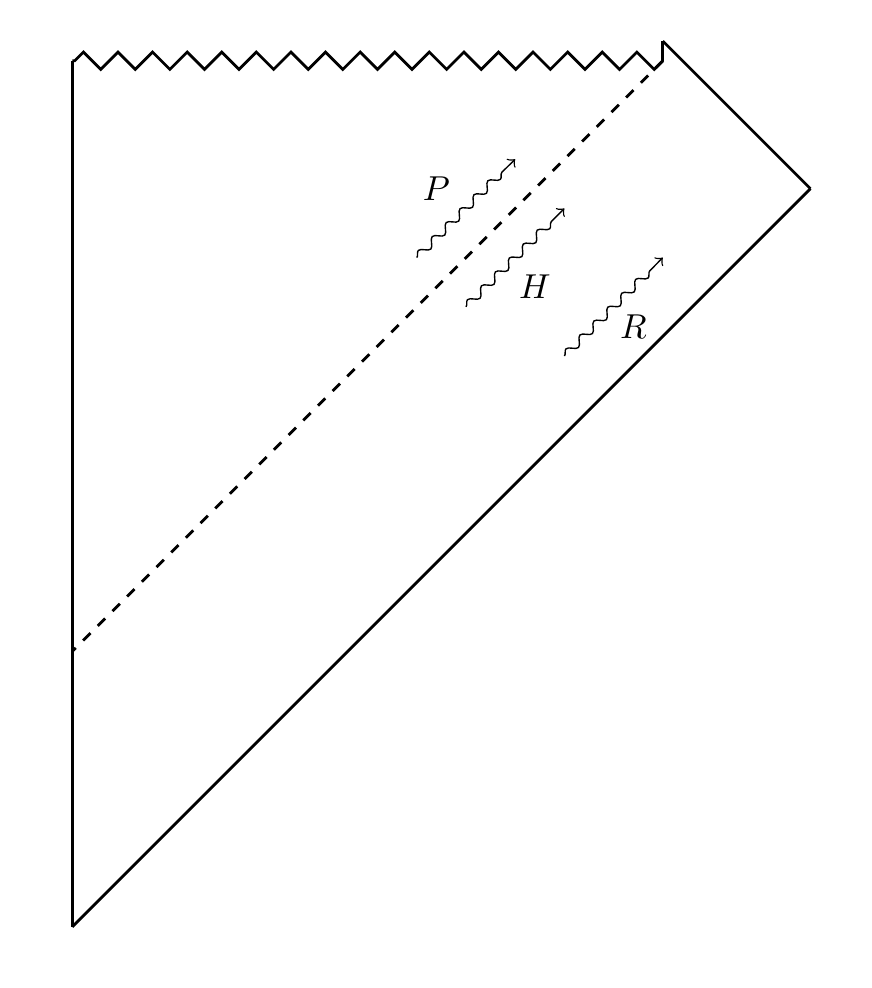}
\caption{Penrose diagram depicting the near-horizon Hawking mode $H$, its behind-the-horizon partner $P$, and the early radiation $R$.\label{fig:Penrose}}
\end{figure}

Recently, Almheiri, Marolf, Polchinski and Sully (AMPS) \cite{AMPS} identified a remarkable conflict between fundamental physical principles. Consider an ``old'' black hole---one that has already emitted more than half of the Hawking quanta---and focus on the emission of the next Hawking photon $H$. The equivalence principle requires that the region near the horizon should look locally like the Minkowski vacuum, requiring that $H$ be strongly entangled with its ``partner mode'' $P$ behind the horizon (see Fig. \ref{fig:Penrose}). However, unitarity requires that $H$ be strongly entangled with the radiation $R$ that has already been emitted. The monogamy of entanglement prohibits $H$ from being maximally entangled with two distinct systems, and locality dictates that $P$ and $H$ are independent.

The essential conflict stated here was already present in Hawking's original work \cite{Hawking_1976}, and was phrased clearly in terms of entanglement by \cite{Mat09, BraPir09}. However, before the work of AMPS, the information paradox could be addressed with black hole complementarity (BHC) \cite{BHC}. In a nutshell, the postulates of BHC simply state that no observer ever witnesses a violation of any physical law, since causality restricts access to all the necessary information. Observers who remain outside the black hole have access to $H$ and $R$ and can thus confirm the unitarity of black hole evaporation, while an infalling observer has access to $H$ and $P$ and can verify the equivalence principle. 

One key innovation of AMPS was to consider the causal patch of an observer who falls into an old black hole. Such an observer would seem to have access to all three ingredients necessary for the paradox. If that is the case, then black hole complementarity is no longer sufficient to resolve the information paradox. However, a closer inspection of the geometrical limitations of a causal patch may reveal deeper issues in need of investigation, such as those proposed in \cite{Ilgin_Yang_2013,Freivogel_2014}. 

In this paper, we analyze another geometric question: can any single observer see the entire sphere behind the horizon? This is a very relevant question because the simplest and most robust version of the paradox requires that the Hawking quantum $H$ actually escapes from the black hole. Due to the angular momentum barrier, Hawking radiation occurs almost exclusively in modes with low angular momentum $\ell$. Furthermore, one encounters subtleties when trying to address the issue of high-$\ell$ modes, which we comment on below. Hence we focus our attention on the s-wave firewall, as this version of the AMPS paradox is both the most fundamental conceptually as well as least ambiguous mathematically. In this model, $H$ should be spread over the entire sphere near the horizon, so its entangled partner mode $P$ is also spread over the entire sphere. Therefore, an observer who cannot see the entire behind-the-horizon sphere will have difficulties recognizing the entanglement between these two modes. Thus, in the context of complementarity, classical considerations are necessary to determine whether any observer can identify the relevant quantum state.

We analyze this question for static black holes in all dimensions, in spacetimes with positive, negative, and zero cosmological constant. For static black holes in asymptotically Minkowski spacetime, an infalling observer cannot receive signals from the entire sphere behind the horizon before hitting the singularity. The most interesting case is 3+1 dimensions, where an observer can see nearly the entire sphere, but with an important caveat: there is a tradeoff between the radial and angular extent of the causal patch, as we describe. In higher dimensions, less than half the sphere fits within one causal patch. 

Adding a negative cosmological constant decreases the region that is causally accessible; for large black holes in asymptotically anti-de Sitter spacetime, in 3+1 and higher dimensions, an infalling observer can only see a small fraction of the horizon sphere, with physical size of order the AdS radius. This result  is potentially important for the AdS version of the firewall paradox \cite{Marolf_Polchinski_2013,AMPSS}, which some consider to be the most robust against the concerns of computation time \cite{HH} and backreaction \cite{HuiYan13}. Since an infalling observer can only see sub-AdS scales near the horizon, the subtle issue of reconstructing these modes from CFT data can play an important role in the firewall paradox \cite{Bousso_etal_2013, Leichenauer_Rosenhaus_2013, Rosenhaus_Rey_2014}. 

Adding a positive cosmological constant increases the angular size of the causal patch. However, we show that once the cosmological constant is large enough to allow an observer to collect information from the entire sphere, the information contained in the Hawking radiation cannot fit within the cosmological horizon. In other words, as the cosmological constant is increased, an infalling observer begins to be able to measure $P$ but loses the ability to measure $R$.

These geometrical results motivate a possible resolution of the firewall paradox: even for an old black hole, some degrees of freedom that are smeared over the entire sphere in the near-horizon zone are entangled with the early radiation, while localized modes in the near-horizon zone are entangled with their partners behind the horizon. This would then avoid an observable conflict between the equivalence principle and unitarity.

There are reasons to think that the AMPS paradox can be reformulated to only refer to modes within a single causal patch. However, existing arguments tend to assume that the geometry allows the measurement of any desired mode. As we endeavor to show, this is often not the case. Causal patch considerations must be taken into account in order to formulate the paradox as cleanly as possible. Our results thus serve as motivation for such a reformulation. 

The organization of this paper is as follows: in Sec. \ref{sec:nonSchwarz} we discuss results for various static black holes, except for Schwarzschild black holes in $3+1$ dimensions, which we treat separately in Sec. \ref{sec:Schwarz}. The reason for this separation is that with the exception of the latter, it is clear that the geometry of the causal patch alone offers an escape from the firewall paradox. In the case of $(3+1)$-dimensional Schwarzschild black holes however, a more detailed analysis is required which occupies the bulk of this work. Then, in Sec. \ref{sec:entropy}, we discuss the consequences for entropy and information in the context of the casual patch considerations in the $(3+1)$-dimensional Schwarzschild background. 

\paragraph{Cases and issues not addressed here.}
BTZ black holes [black holes in (2+1)-dimensional AdS spacetime], named for Ba\~nados, Teitelboim, and Zanelli, are an exception: in this case an infalling observer can collect information from the entire sphere behind the horizon. The physics of black holes in 2+1 dimensions is rather different than in higher dimensions -- for example, there are no black holes in asymptotically flat space in 2+1 dimensions. We leave them aside for the purpose of this analysis, but it may be interesting to further consider this case. 

We do not treat rotating black holes in this paper. In this case, there is no spherical symmetry, so it is less obvious which sphere must be contained within the causal patch in order to formulate the paradox. Additionally, due to the presence of a nearly null inner horizon, light rays may be able to travel farther before hitting the singularity. We leave this analysis for future work.

 An additional issue concerns black hole mining. AMPS argued that the high-$\ell$ modes must also be entangled with the early radiation. Their arguments involved ``mining'' black holes: inserting a device such as a string that collects radiation from deep in the zone and transports it to the exterior.  Brown \cite{Brown_2012} derived a number of interesting constraints on black hole mining, including the constraint that the mining equipment must be smaller than the local thermal wavelength of the Hawking radiation. Furthermore, in order to successfully extract energy and information from the black hole, the mining device must be nearly static. But clearly the presence of such a device can disrupt the entanglement between the relevant Hawking mode that is mined and its partner behind the horizon. The process of deploying this mining device may also disrupt the entanglement between the late quanta and the early radiation. We regard mining as an interesting direction for future work. Here, we restrict our analysis to unmined black holes, where the outgoing radiation is almost exclusively in the modes with low angular momentum on the sphere.

\section{Static black holes in higher dimensions}\label{sec:nonSchwarz}
In this section, we consider arbitrary dimensional static black holes in spacetimes with positive, negative, and zero cosmological constant. We postpone a detailed discussion of the critical $(3+1)$-dimensional static black hole to the next section, as the geometry of the causal patch and its implications for the firewall discussion are more subtle in this case.

\subsection{Black holes in asymptotically Minkowski spacetime}
To address the question of how much of the sphere an infalling observer can see, we need to calculate the maximum angle a light ray can travel between the horizon and the singularity. For static black holes in $D>3$ spacetime dimensions, the metric is
\be
\dd s^2 = - f(r) \dd t^2 + {\dr^2 \over f(r)} + r^2 \dd\Omega_{D-2}^2
\ee
where
\begin{align}
f(r)=\left[1-\left(\frac{r_-}{r}\right)^{D-3}\right]\left[1-\left(\frac{r_+}{r}\right)^{D-3}\right]
\end{align}
with
\begin{align}
r_\pm=\frac12 \left(r_s \pm \sqrt{r_s^2-4 r_Q^2} \right)
\end{align}
Here $r_+$ and $r_-$ are the radii of the outer and inner horizons, respectively; the parameter $r_Q$ is determined by the charge of the black hole, and is given by $r_Q^2=Q^2G/\left(4\pi\epsilon_0c^4\right)$. For uncharged black holes, $r_Q=0$ and the above reduces to the Schwarzschild solution ($r_-\rightarrow0$, $r_+\rightarrow r_s$) with Schwarzschild radius $r_s$. For the Reissner-Nordstrom solution ($Q^2>0$), the inner horizon is believed to be unstable to perturbations, so the natural question is how far light rays can travel between the outer horizon and inner horizon in the angular direction. 

Inside the outer horizon, the $r$ and $t$ coordinates switch roles, such that $r$ is temporal and $t$ is spatial. Hence to move the maximum distance along the sphere, the ray should not move in the $t$ direction. Therefore the null ray that travels the maximal angle satisfies
\be
r^2 \dd\theta^2 = - {\dr^2 \over f(r)}
\ee
and the angle is given by
\be
\Delta \theta = \int_{r_-}^{r_+} {\dr \over r \sqrt{-f(r)}}=\frac{\pi}{D-3}\label{eq:delTheta}
\ee
Thus for higher-dimensional black holes, it is impossible for a single observer to see the entire horizon, and therefore such an observer will have difficulty identifying the quantum state necessary to formulate the paradox in the global framework.\footnote{Although existing versions of the paradox rely on a global picture, it may be possible to formulate a local version of the paradox, which might allow one to evade such concerns.} For the limiting case $D=4$, there is at most just enough time for the information to be collected at a point, but no time for it to be processed. If the same property holds for all black holes, it suggests a principle: a freely falling observer cannot access the entire horizon sphere, and therefore cannot measure modes of definite angular momentum.

\subsection{Black holes in de Sitter}
One can ask about the effect of a nonzero cosmological constant on the above calculation. In this section we show that introducing a positive cosmological constant increases $\Delta \theta$, allowing the infalling observer to fit the entire infalling sphere inside her causal diamond. However, at the same time the cosmological horizon moves closer to the black hole. We find that by the time the cosmological constant is large enough to allow the infalling observer to see the entire sphere, the cosmological horizon is too small to allow for the early radiation to be collected.

\subsubsection{3+1 dimensions}
Introducing a positive cosmological constant will change the metric so that now
\be
f(r) = 1 - {M \over r} - \frac{r^2}{b^2}\label{eq:fdS}
\ee
where $M$ is the black hole mass and $b^2\equiv3/\Lambda$. We want to know how this affects the angle computed above -- will putting black holes in de Sitter space allow the infalling observer to see the entire horizon sphere?

Using again \eqref{eq:delTheta} for the angle, we get
\be
\Delta \theta = \int_0^{r_H} {\dr \over r \sqrt{ -1 + {M \over r} + \frac{r^2}{b^2}}} = b\int_0^{r_H} {\dr \over \sqrt{r} \sqrt{(r - r_1) (r - r_2) (r - r_3)}}
\ee
where the $r_i$ are the three roots of the equation $f(r)=0$. If we assume that $M<M_c\equiv \frac{2b}{3 \sqrt{3}}$, these three roots are the black hole horizon $r_H$, the cosmological horizon $r_c$, and a third negative root $r_3 = - r_H - r_c$.  Defining a dimensionless variable $u = r/r_H$ and rearranging gives
\be
\Delta \theta = b\int_0^1\frac{\mathrm{d}u}{\sqrt{u - u^2}\sqrt{{r_c(r_c+ r_H)} - r_H^2(u + u^2)}}
\ee
Note that in the limit that the dS radius is much bigger than the black hole, $r_c \approx b$ and the second factor approaches 1, giving the flat space result.

We would like to approximate the formula for $r_H \ll r_c$. First we use that the product of the three roots is $\prod_i r_i = -M b^2$, so
\be
r_c (r_c + r_H) = {M b^2 \over r_H}
\ee
so that 
\be
\Delta \theta = \int_0^1 \frac{\mathrm{d}u}{\sqrt{u - u^2}\sqrt{{M \over r_H} - {r_H^2 \over b^2}(u + u^2)}}
\ee
Now, perturbatively solving \eqref{eq:fdS} for $r_H$ and taking the limit where $r_H\approx M$ yields
\be
{M \over r_H} = 1 - {M^2 \over b^2} + \dots
\ee
so that finally the integral of interest is
\be
\Delta \theta \approx \int_0^1 \frac{\mathrm{d}u}{\sqrt{u - u^2}\sqrt{1 - \frac{r_H^2}{b^2}(1 + u + u^2)}}\approx\pi+\frac{15\pi}{16}\frac{r_H^2}{b^2}
\ee
A nice way to summarize this result is to write it in terms of the entropy of the two horizons:
\be
\Delta \theta = \pi + \frac{15\pi}{16}{S_{BH} \over S_{dS}}\label{eq:thetadS}
\ee
This shows that in principle an observer inside has access to the entire horizon sphere in some location. Now suppose that we want to collect the information at least a Planck distance from the singularity -- then instead of integrating all the way to $r = 0$ we should integrate to the location $r=r_p$ where
\be
\int_0^{r_P} {\mathrm{d}r \over \sqrt{-f(r)}} \approx  \int_0^{r_P}{\mathrm{d}r \sqrt{r/M}} = \frac23\frac{r_P^{3/2}}{\sqrt{M}}\equiv l_P
\ee
so that $r_P=\left(\frac32\right)^{2/3} l_P^{2/3} M^{1/3}$, giving a lower cutoff on the $u$ integral of $u_P=r_P/r_H\approx \left(\frac32\right)^{2/3} \left(\frac{l_P} {M}\right)^{2/3}$, where we used that $r_H\approx M$. This corrects the angle by about
\be
\int_0^{u_P} {\mathrm{d}u \over \sqrt u} =2 \sqrt{u_P}\approx 12^{1/3} \pi^{1/6}S_{BH}^{-1/6}
\ee
So overall, the angular distance that light can travel behind the horizon of a Schwarzschild black hole in de Sitter space before reaching regions of Planckian curvature is
\be
\Delta \theta = \pi + \frac{15\pi}{16}{S_{BH} \over S_{dS}} - 12^{1/3} \pi^{1/6}S_{BH}^{-1/6}
\ee
and, at this level of analysis, we can see the entire horizon as long as 
\be
S_{dS} < S_{BH}^{7/6} 
\label{ineq}
\ee
where we have neglected order 1 factors. However, the amount of information that can be stored inside the horizon in any ordinary system is \cite{Bousso_Freivogel_Leichenauer_2010}
\be
S_R < S_{dS}^{3/4}\label{sbound}
\ee
Since we need to be able to collect a number of bits comparable to the black hole entropy, $S_R \sim S_{BH}$. Therefore, the combined constraints on the size of the cosmological horizon give
\be
S_{BH}^{4/3} < S_{dS} < S_{BH}^{7/6}~.
\ee
But since $S_{dS}$ is larger than 1, $S_{BH}^{4/3} > S_{BH}^{7/6}$, so the combined inequality cannot be satisfied.

Therefore, whenever the cosmological constant is large enough to allow the infalling observer to see the partner modes behind the horizon, the AMPS paradox cannot be constructed for another reason: the Hawking radiation will not fit inside the cosmological horizon.

\subsubsection{Higher dimensions} 
For dS black holes in arbitrary dimensions, \eqref{sbound} becomes $S_R<S_{dS}^{(D-1)/D}$. This means that for large black holes whose radiation can be collected within the causal patch, the cosmological horizon $b$ is much larger than the black hole horizon $r_H$. In this limit, the higher-order corrections to the flat space result $\Delta \theta = \frac{\pi}{D-3}$ are small, so they do not change the conclusion that the observer is missing an order 1 fraction of the sphere. Therefore, as long as $S_R$ fits inside the cosmological horizon, the infalling observer cannot see the entire horizon sphere. 

\subsection{Black holes in anti-de Sitter}
For AdS-Schwarzschild black holes, the result is very interesting. In this case we will work in general $D$-dimensional spacetime, where $D\geq 4$. The metric function for an AdS black hole is given by
\be
f(r) = 1 + {r^2 \over b^2} - {R_S^{D-3} \over r^{D-3}}
\ee
where for AdS we have $b^2\equiv-3/\Lambda>0$. The relevant integral is
\be
\Delta \theta = \int_0^{r_H} {\mathrm{d}r \over r \sqrt{-f(r)} }~.
\ee
For a large black hole with  horizon radius much larger than the AdS radius, it is important to ask how large the part of the horizon is that fits inside one causal patch: is it many AdS radii, or not? Taking the large black hole limit, we get
\begin{align}
\Delta \theta &\approx \int_0^{r_H} {\mathrm{d}r \over r\sqrt{{R_S^{D-3}\over r^{D-3}} - {r^2\over b^2}}} \\
&= \int_0^{r_H} {\mathrm{d}r \over R_S^{D-3\over 2}r^{5-D\over 2}\sqrt{1 - {r^{D-1}\over R_S^{D-3}b^2}} }
\end{align}
In the $b^2\ll r_H^2$ limit we can use that $r_H^{D-1} \approx R_S^{D-3} b^2$ and change variables to get the dependence on parameters outside the integral, giving
\be
\Delta \theta = {b \over r_H} \int_0^1 \mathrm{d}u {u^{D-5\over 2} \over \sqrt{1-u^{D-1}}} \sim {b \over r_H}
\ee
where the integral can be evaluated exactly to give an $O(1)$ number for $D=4$ which is monotonically decreasing with increasing $D$.
This shows that for a big black hole in AdS, only a small fraction of the horizon fits inside the causal patch of an infalling observer. The corresponding physical length along the horizon that fits in one causal patch is
\be
\Delta x \sim b.
\ee
We can conclude that an observer falling into a large AdS-Schwarzschild black hole in a $D$-dimensional spacetime has access to only a small part of the horizon, with physical size of order one AdS radius.

This fact may have important consequences for the AdS/CFT version of the firewall argument \cite{Marolf_Polchinski_2013}. Existing techniques for mapping bulk to boundary encounter interesting complications when applied to fields localized to less than one AdS radius in the near-horizon region \cite{Kabat:2011rz, Bousso_etal_2013}. 
It is very intriguing that the arguments for a firewall in AdS black holes must focus on phenomena within a single AdS radius. It is precisely in this regime that the AdS/CFT duality is not well understood, and there are obstacles to reconstructing the bulk physics from the CFT.

\section{Black holes in 3+1 dimensions}\label{sec:Schwarz}
As indicated by \eqref{eq:delTheta}, for (3+1)-dimensional black holes in asymptotically Minkowski spacetime, an infalling observer can see the entire sphere just as she hits the singularity. This case calls for a more detailed analysis of the causal patch.

A full analysis requires the inclusion of both interior and exterior s-wave partners, and thus we must identify a spacelike slice that crosses the horizon of the black hole. We want to know about the physics of observers who fall in to the black hole from infinity. The Gullstrand-Painlev\'{e} (GP, a.k.a. ``rain-frame'') coordinates are ideally suited for such purposes: the GP time variable $T$ is the proper time along the worldline of observers falling into the black hole, starting from rest at infinity. The slices of constant $T$ are thus orthogonal to such observers, and have the additional appeal of being spatially flat. Therefore, analyzing the entanglement in this frame is directly relevant to the question of whether an infalling observer detects any violation of the equivalence principle, as the geometrical properties of the GP coordinates precisely reflect the causal evolution along an infalling trajectory.

\begin{figure}[h!]
\centering
\includegraphics[width=0.44\textwidth]{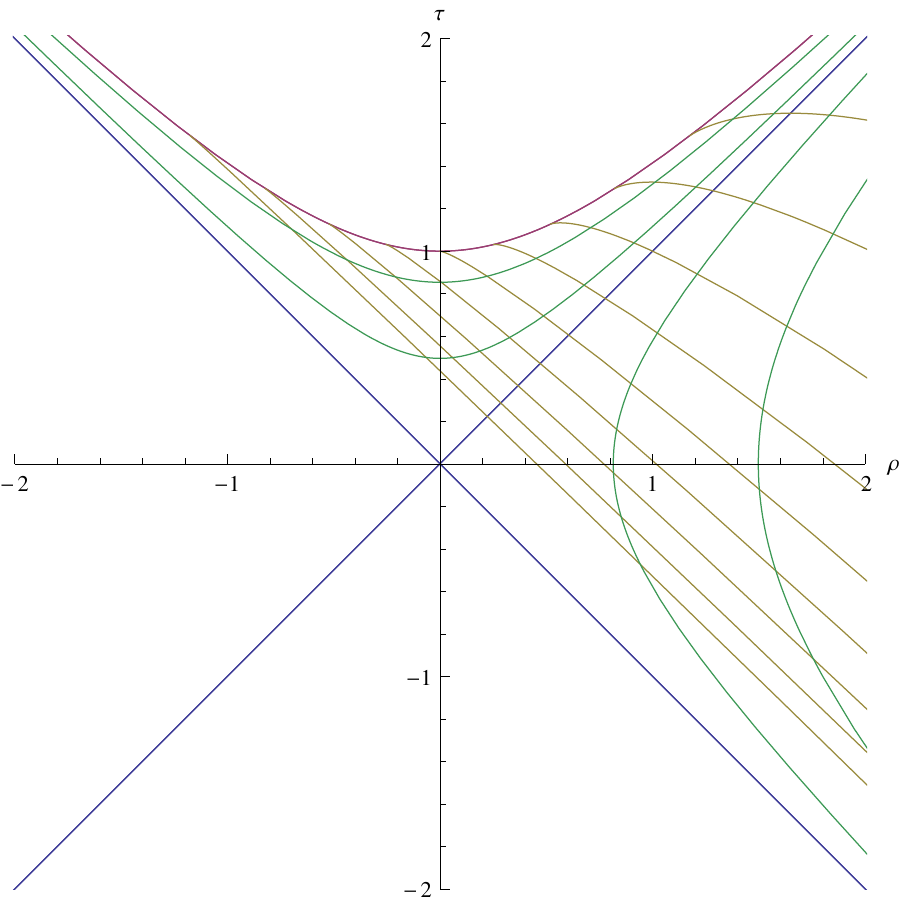}
\caption{(color online). Schwarzschild black hole in Gullstrand-Painlev\'{e} coordinates, with singularity at $r=0$ (top-most curved line, red in online color version), showing constant $r$ slices (curved lines, green online), and constant $T$ slices (slanted lines, yellow online). The vertical and horizontal axes are Kruskal-Szekeres time and radius, respectively, while the Schwarzschild radius has been set to $r=1$.\label{fig1}}
\end{figure}

The GP coordinates are  defined as follows \cite{Martel_Poisson_2000}:
Beginning with the Schwarzschild metric, define a new coordinate
\be
T=t+r_s\left(2\sqrt{\frac{r}{r_s}}+\ln\left|\frac{\sqrt{\frac{r}{r_s}}-1}{\sqrt{\frac{r}{r_s}}+1}\right|\right)\label{eq:TGP}
\ee
called the Gullstrand-Painlev\'{e} time, with which the metric may be rewritten
\be
\dd s^2=-f\dd T^2+2\sqrt{\frac{r_s}{r}}\dd T\dr+\dr^2+r^2\dd\Omega^2
\ee
which has the appeal of being regular at $r=r_s$. See Fig. \ref{fig1} for a depiction of the constant $T$ slices.

We want to determine the causal structure, so we need the equation for null geodesics in these coordinates. The conserved quantities for the GP metric are
\begin{align}
E&=f\dot T-\sqrt{\frac{r_s}{r}}\dot r\label{eq:energyGP}\\
\ell&=r^2\dot\theta
\end{align}
where the dot denotes differentiation with respect to some affine parameter. By using the second of these to replace $\dot\theta$ in the null geodesic equation $\dd s^2=0$, and using the resulting expression for $\dot T$ in \eqref{eq:energyGP}, one obtains a third conservation expression:
\be
E^2=\dot r^2+\frac{f}{r^2}\ell^2
\ee
which we may use to eliminate the affine parameter and obtain an expression for the angular distance traversed by an arbitrary null geodesic:
\be
\frac{\dot\theta}{\dot r}=\frac{\mathrm{d}\theta}{\mathrm{d}{r}}\implies\Delta\theta=\int_0^{r'}\frac{\pm\mathrm{d}r}{\sqrt{\epsilon^2 r^4+ r^2f}}\label{eq:theta}
\ee
where $\epsilon\equiv E/\ell$, and the $\pm$ sign selects the polar direction in which the null ray travels. Note the fundamental difference between this expression and \eqref{eq:delTheta}: our null rays are no longer constrained to move along constant Schwarzschild $t$-slices in the black hole interior.

Similarly, we obtain an expression for the Gullstrand-Painlev\'{e} time difference corresponding to \eqref{eq:theta}:
\be
\frac{\dot T}{\dot r}=\frac{\mathrm{d}T}{\mathrm{d}{r}}\implies\Delta T=\int_0^{r'}\frac{1}{f}\left(\sqrt{\frac{r_s}{r}}\pm\frac{\epsilon r}{\sqrt{\epsilon^2 r^2-f}}\right)\mathrm{d}r\label{eq:T}
\ee
Henceforth we will absorb the $\pm$ sign in our expressions for $\Delta T$ into $\epsilon$ by allowing the latter to take negative values.

\begin{figure}[h!]
\centering
\includegraphics[width=0.44\textwidth]{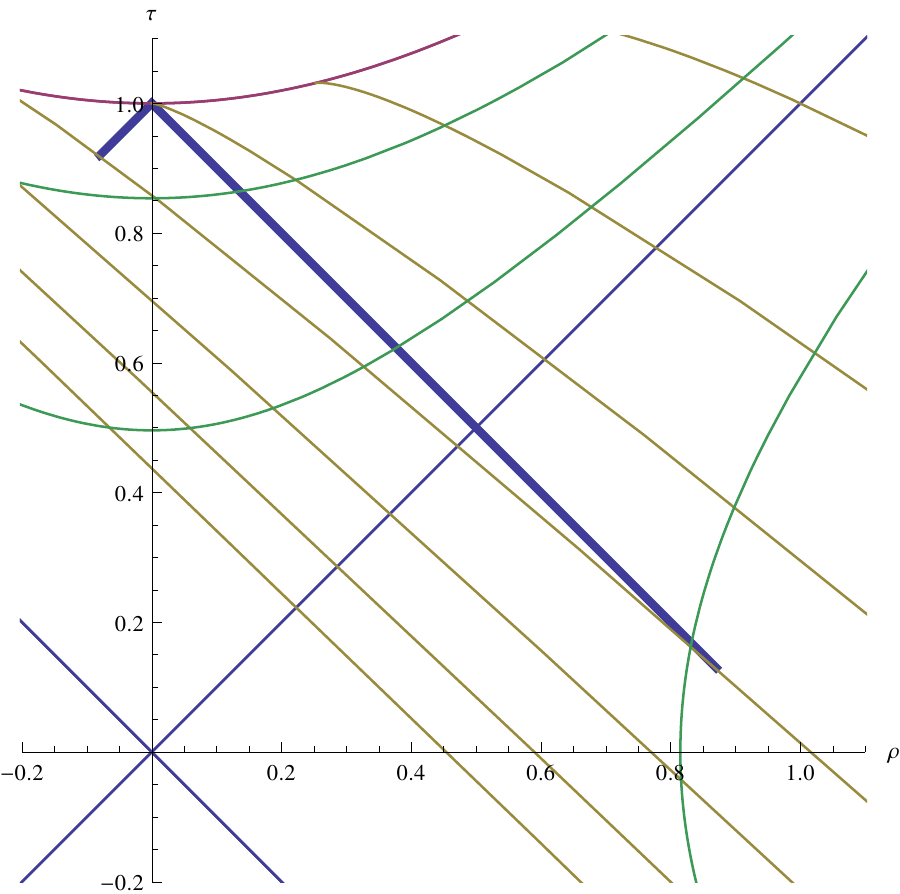}
\caption{(color online). Past light cone (bold blue) of an observer hovering just above the singularity at $(t,r)\approx(0,0)$. The interior and exterior radial null rays (left and right cone sides, respectively) intersect the $T$-slice at $r_{\epsilon\rightarrow\infty}$, $r_{\epsilon\rightarrow -\infty}$, respectively. The geometry of the patch is determined by evaluating $\Delta\theta$ along the $T$-slice for the null rays between these two radial extremes.\label{fig2}}
\end{figure}

Now we would like to determine which part of the constant time surface fits within a single causal patch.   We fix a single observer, who determines the causal patch, just above the singularity at Schwarzschild time $t=0$, at the north pole of the sphere, $\theta = 0$. This observer will collect measurements transmitted to her from an infalling distributed measuring device -- say, a ring of probes spread around the horizon. At some specified GP time $T$, the probes will perform a measurement of the interior s-wave and transmit this information to the observer to be collected for analysis. The intersection of the observer's past light cone with this $T$-slice determines the causal patch under consideration (see Fig. \ref{fig2}).

The Schwarzschild time of the observer ($t=0$) intersects this $T$-slice at $r=r_0$. We wish to know the geometry of this causal patch as a function of the choice of $T$ (equivalent to considering observers who fall in at different Schwarzschild times), which requires numerically evaluating \eqref{eq:theta} along the $T$-slice.

\begin{figure}[h!]
\centering
\begin{tabular}{cc}
\begin{subfigure}{0.49\textwidth}
	\includegraphics[width=0.9\textwidth]{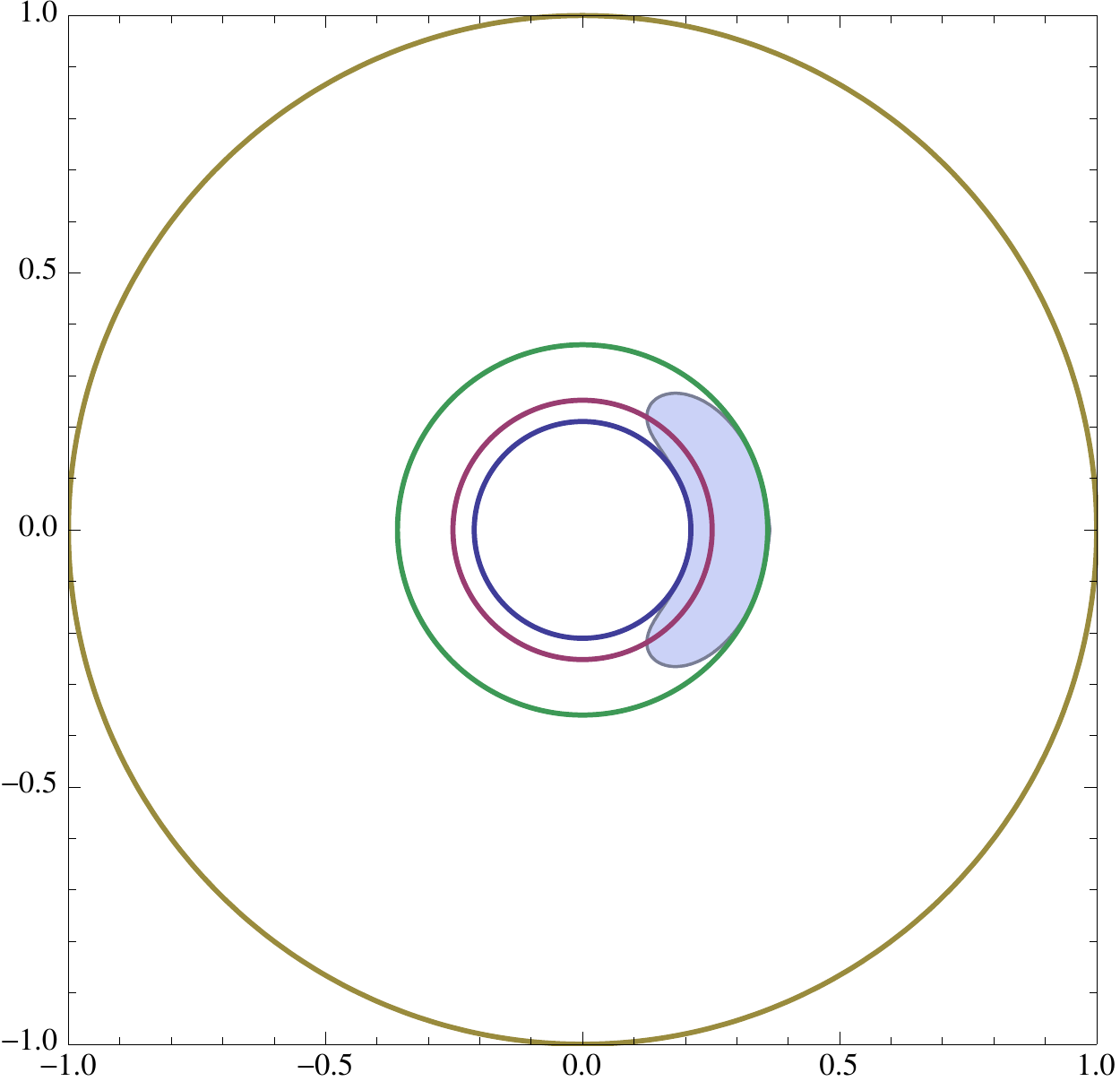}
	\caption{$|\Delta T|=0.1$}
\end{subfigure} & 
\begin{subfigure}{0.49\textwidth}
	\includegraphics[width=0.9\textwidth]{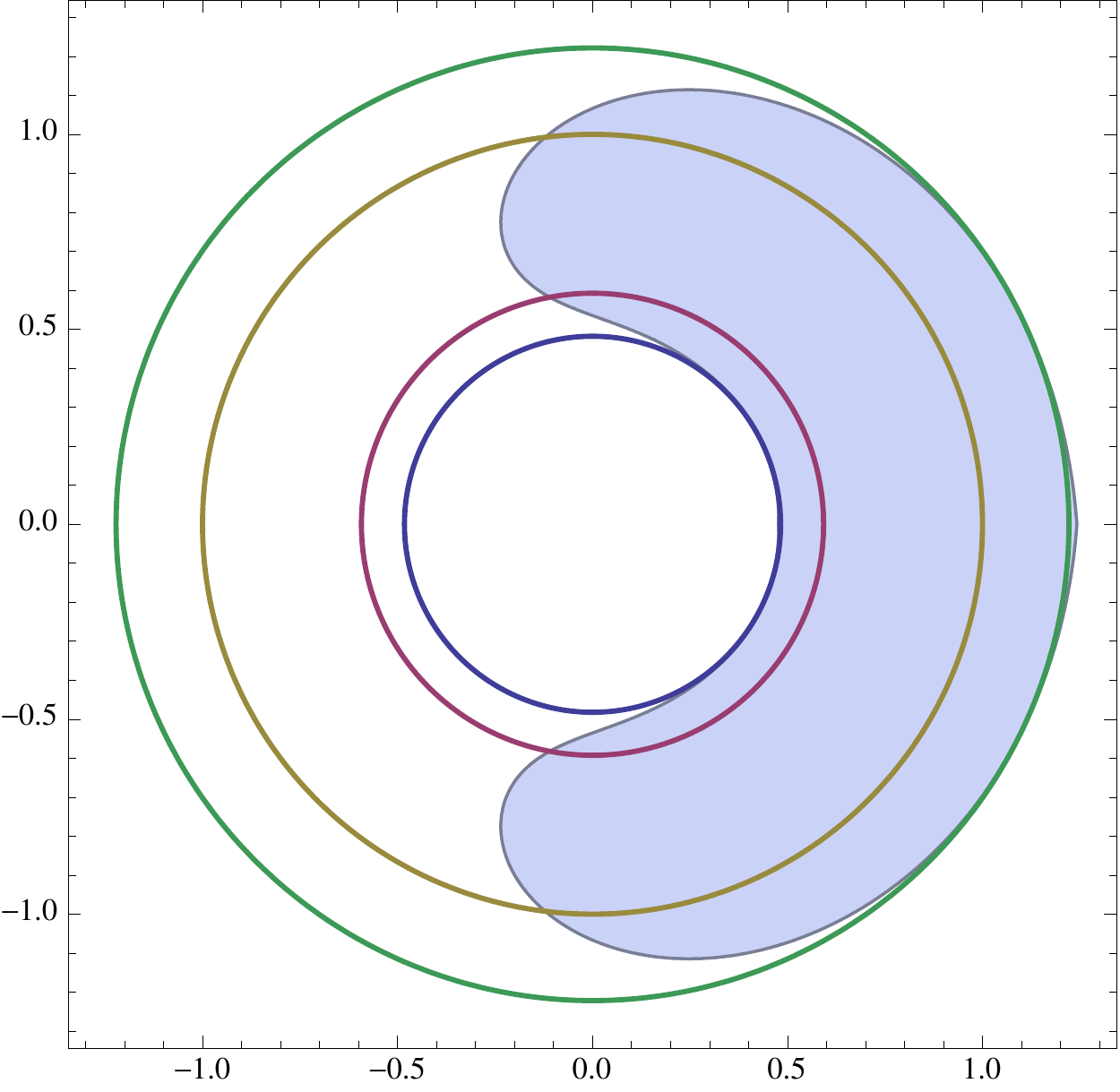}
	\caption{$|\Delta T|=0.5$}
\end{subfigure} \\\\
\begin{subfigure}{0.49\textwidth}
	\includegraphics[width=0.9\textwidth]{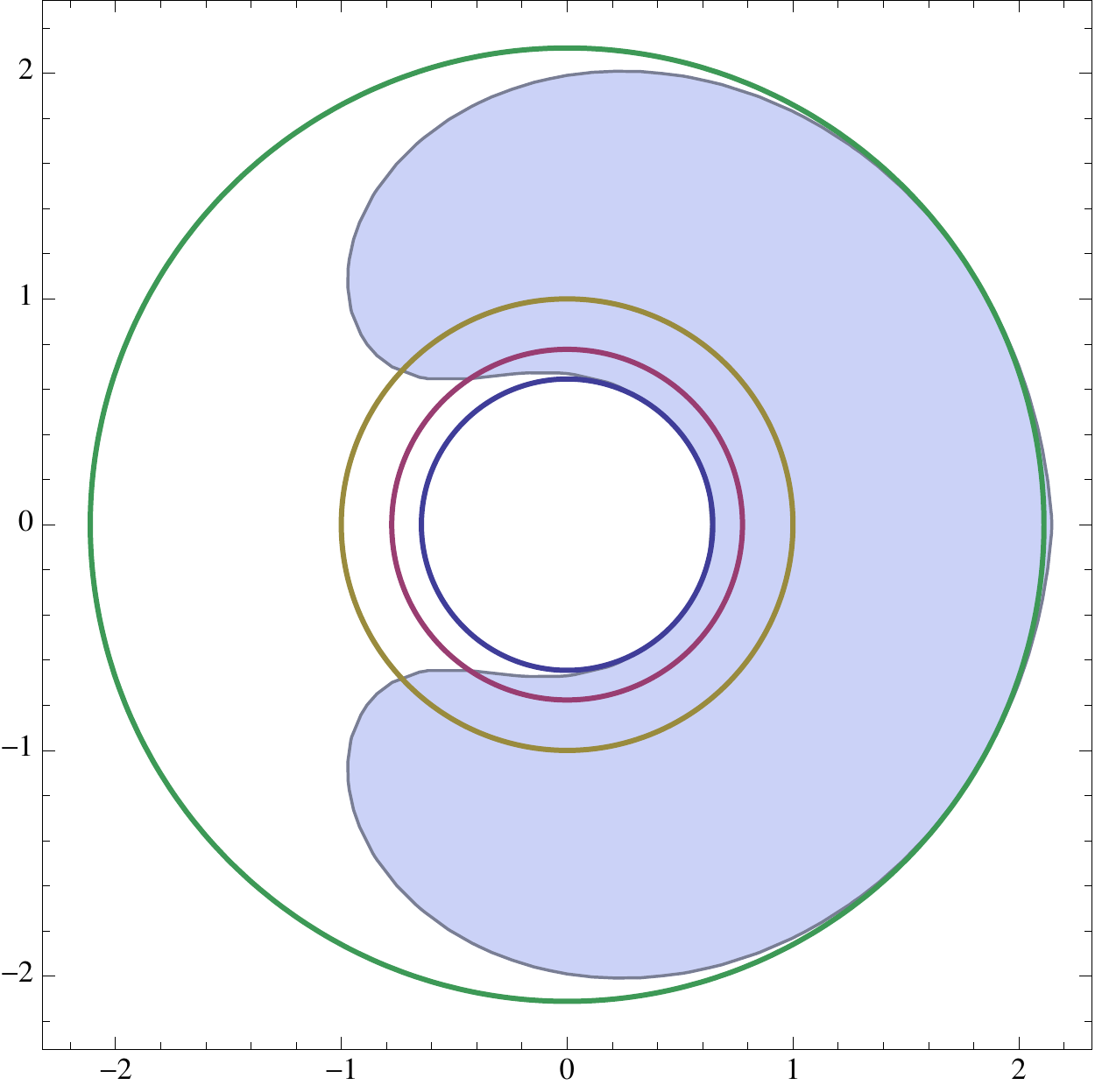}
	\caption{$|\Delta T|=1$}
\end{subfigure} &
\begin{subfigure}{0.49\textwidth}
	\includegraphics[width=0.9\textwidth]{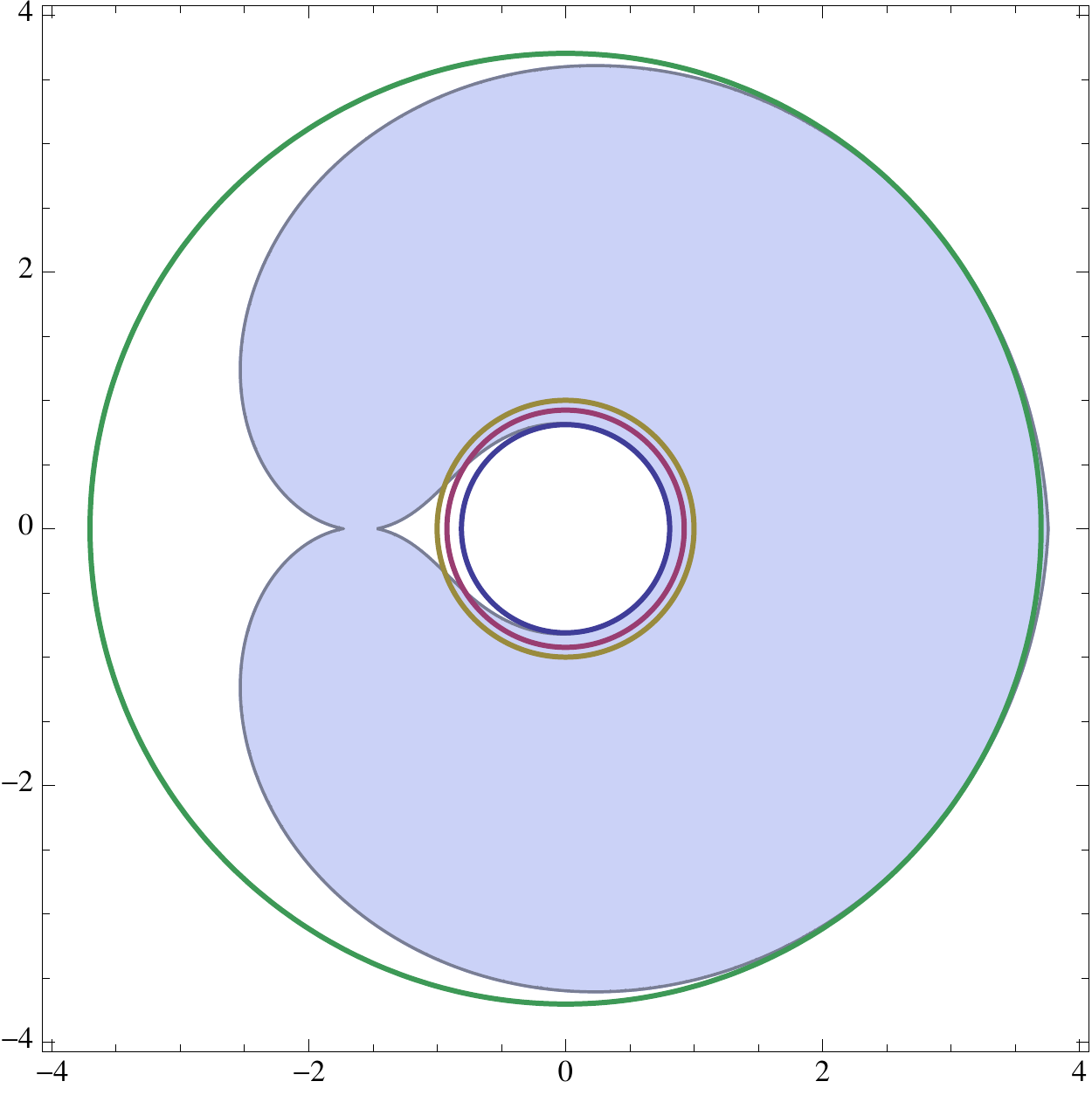}
	\caption{$|\Delta T|=2$}
\end{subfigure} \\
\end{tabular}
\caption{(color online). Causal patch geometry for several values of $\Delta T$. The shaded region depicts the portion of the spacelike $T$-slice [as $r(\theta)$] visible to the observer. The concentric rings show the horizon $r_s=1$ (outermost ring in (a), next-to-outermost in (b-d); yellow online), $r_{\epsilon\rightarrow -\infty}$ (outermost ring in (b-d), next-to-outermost in (a); green online), $r_0$ (next-to-innermost ring, red online), and $r_{\epsilon\rightarrow\infty}$ (innermost ring, blue online). (Note that the axes are rescaled between images). Increasing $|\Delta T|$ corresponds to selecting a $T$-slice closer to the past horizon in Fig. \ref{fig2}.\label{fig:exclusionregions}}
\end{figure}

To perform this evaluation requires specification of $\epsilon$. For each point in the causal patch, there intersects in principle an infinite number of possible null rays, parameterized by $\epsilon$, only one of which will have the correct trajectory to be collected by the observer. Furthermore, this value of $\epsilon$ is dependent on the upper limit of integration, i.e. on the $r$-position along the $T$-slice: $\epsilon=0$ corresponds to $\ell\rightarrow\infty$, for which \eqref{eq:theta} reduces to \eqref{eq:delTheta}, while $\epsilon\rightarrow\pm\infty$ corresponds to radial null rays with $\ell=0$, whose intersections with the $T$-slice give the minimal (at $r=r_{\epsilon\rightarrow\infty}$) and maximal ($r=r_{\epsilon\rightarrow -\infty}$) radii of the casual patch. 

The distance between the observer and our chosen $T$-slice, denoted $T_*$, is given by $\Delta T=T_*-T(r=0,t=0)=T_*$. Thus we may numerically obtain the values of $\epsilon$ for radii along $T=T_*$ by finding the root of $T_*-\Delta T(\epsilon)$, where $\Delta T(\epsilon)$ is given by \eqref{eq:T}, with $\epsilon$ as the free parameter. With these values of $\epsilon$ in hand, we may proceed to the numerical evaluation of $\eqref{eq:theta}$. Results are shown in Fig. \ref{fig:exclusionregions}.

As $|\Delta T|$ is increased, the observer sees less of the interior and more of the exterior of the black hole. This is consistent with an inspection of the geometry in Fig. \ref{fig2}: as $T_*$ becomes more and more negative, $r_{\epsilon\rightarrow\infty}$ approaches the horizon radius, while $r_{\epsilon\rightarrow -\infty}$ increases without bound; conversely, as $T_*$ approaches $t=0$, both $r_{\epsilon\rightarrow\infty}$ and $r_{\epsilon\rightarrow -\infty}$ shrink, allowing the observer to see more of the black hole interior at the cost of her external view.

In order to try to fit all the ingredients necessary for the firewall paradox inside a single causal patch, we wish to examine a causal patch that contains both an outgoing Hawking quantum and its interior partner mode. Hence for our purposes, the regime of interest is when $|\Delta T|$ becomes large, which allows the observer to maximize both her internal and external angular visibility, and hence affords the best chance of measuring both an outgoing s-wave and its entangled interior partner. However, as pointed out in \cite{Ilgin_Yang_2013}, the wavelength of the interior mode may pose some difficulty to fitting it inside such a patch. In particular, because of the aforementioned trade-off between angular and radial depth visibility, it may not be possible to keep the wavelength of the interior mode above the Planck scale while effecting sufficient angular resolution.

\begin{figure}[h!]
\centering
\begin{tabular}{cc}
\begin{subfigure}{0.49\textwidth}
	\includegraphics[width=0.9\textwidth]{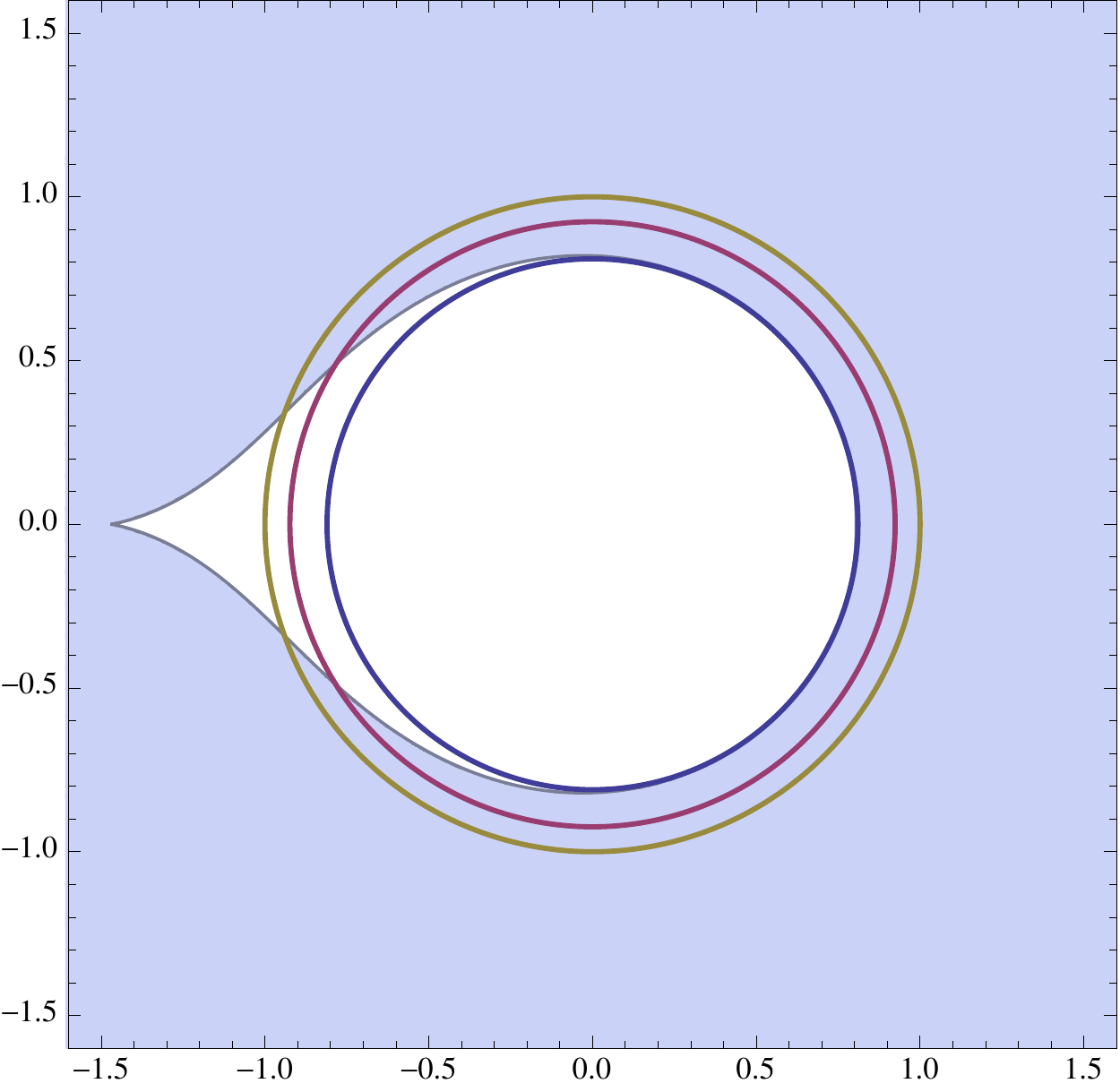}
	\caption{$|\Delta T|=2$}
\end{subfigure} & 
\begin{subfigure}{0.49\textwidth}
	\includegraphics[width=0.9\textwidth]{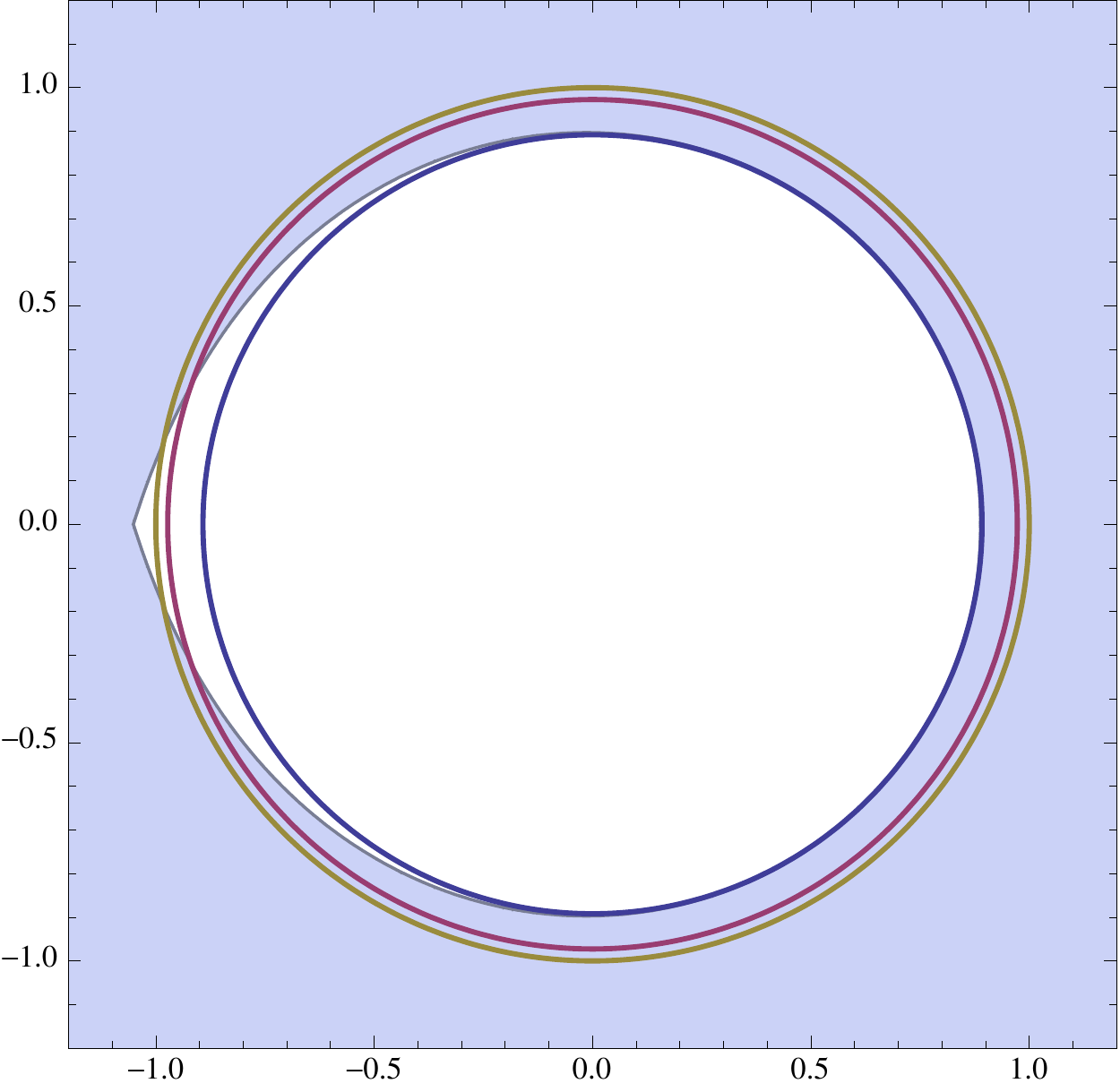}
	\caption{$|\Delta T|=3$}
\end{subfigure} \\
\end{tabular}
\caption{Rain in the rain frame: close-up of exclusion regions. The pointed end of the raindrop diminishes, and the droplet approaches a circular region with radius $r\rightarrow r_s$, in the limit of large $\Delta T$.\label{fig:rain}}
\end{figure}

For $|\Delta T|$ sufficiently large to close the exterior visibility region, the exclusion region resembles a raindrop (see Fig. \ref{fig:rain}). In the limit of large $|\Delta T|$, $r_{\epsilon\rightarrow\infty}$ approaches $r_s$, and the radial depth available to interior s-wave modes vanishes. Since the energy is $\sim \lambda^{-1}$, this places a lower limit on the energy of the measurable modes, namely $E\gtrsim (r_s-r_{\epsilon\rightarrow\infty})^{-1}$.

Although an analytical expression for the droplet geometry is not available, it is possible to obtain an approximation in the large $|\Delta T|$ limit, where the droplet begins to look like that in Fig. \ref{fig:rain} for $|\Delta T|=3$. By approximating Eqs. \eqref{eq:theta} and \eqref{eq:T} in the small-$\epsilon$ limit, we find
\begin{empheq}[box=\widefbox]{align}
\Delta\theta&\approx\pi-2\sqrt{1-r+\ep^2}\label{eq:thetaApprox}\\
\Delta T&\approx2\sqrt{r}+2\nlog\left(\frac{1-\sqrt{r}}{\ep+\sqrt{1-r+\ep^2}}\right)+2\nlog\left(\ep+\sqrt{1-\ep^2}\right)\label{eq:Tapprox}
\end{empheq}
The derivation of these expressions is detailed in the Appendix. Note that $\Delta T<0$ (consistent with an infalling observer, since we integrated outwards from the singularity $r=0$).

\begin{figure}[h!]
\centering
\begin{tabular}{cc}
\begin{subfigure}{0.49\textwidth}
\includegraphics[width=0.9\textwidth]{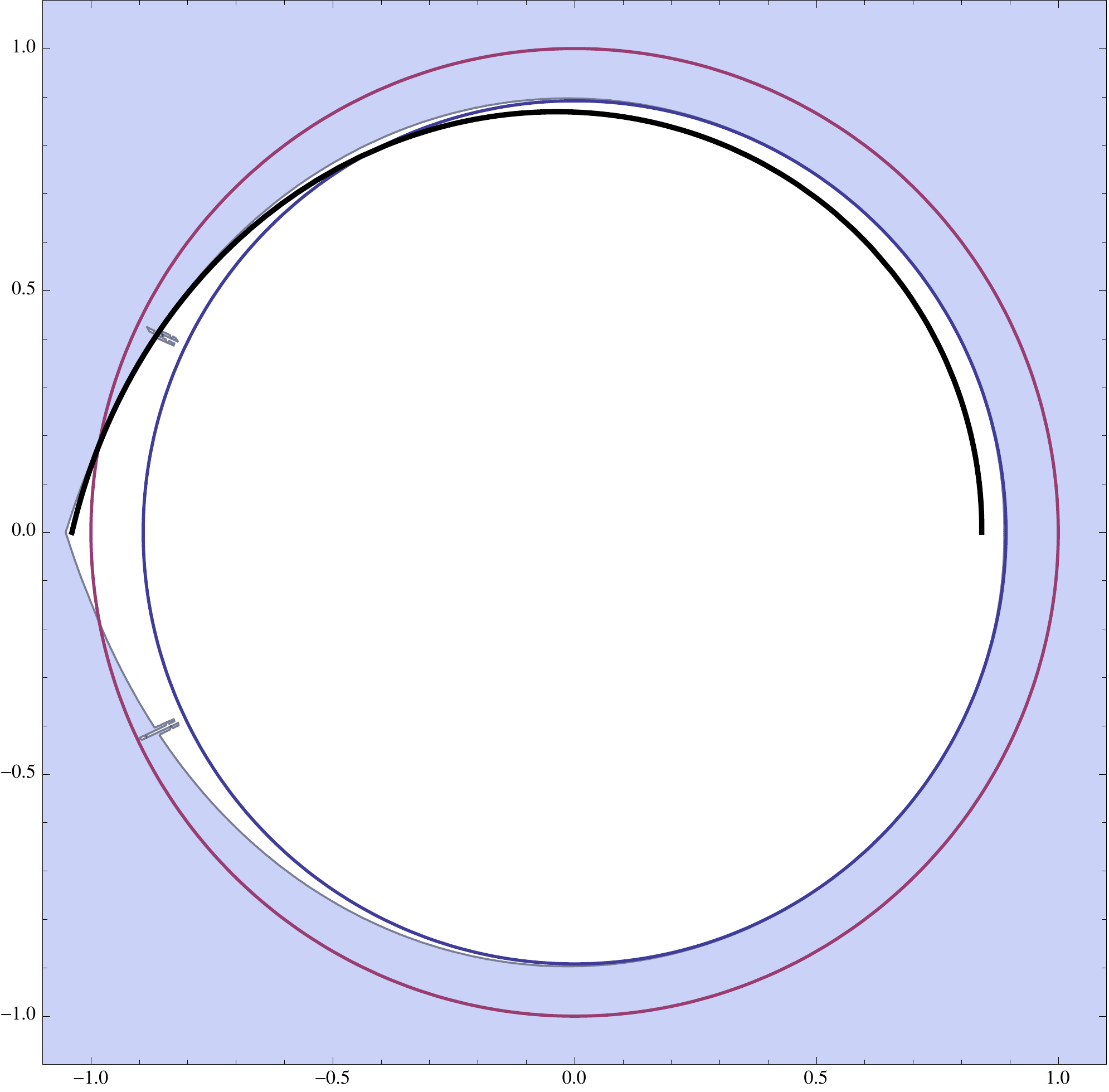}
\end{subfigure}
\begin{subfigure}{0.49\textwidth}
\includegraphics[width=0.9\textwidth]{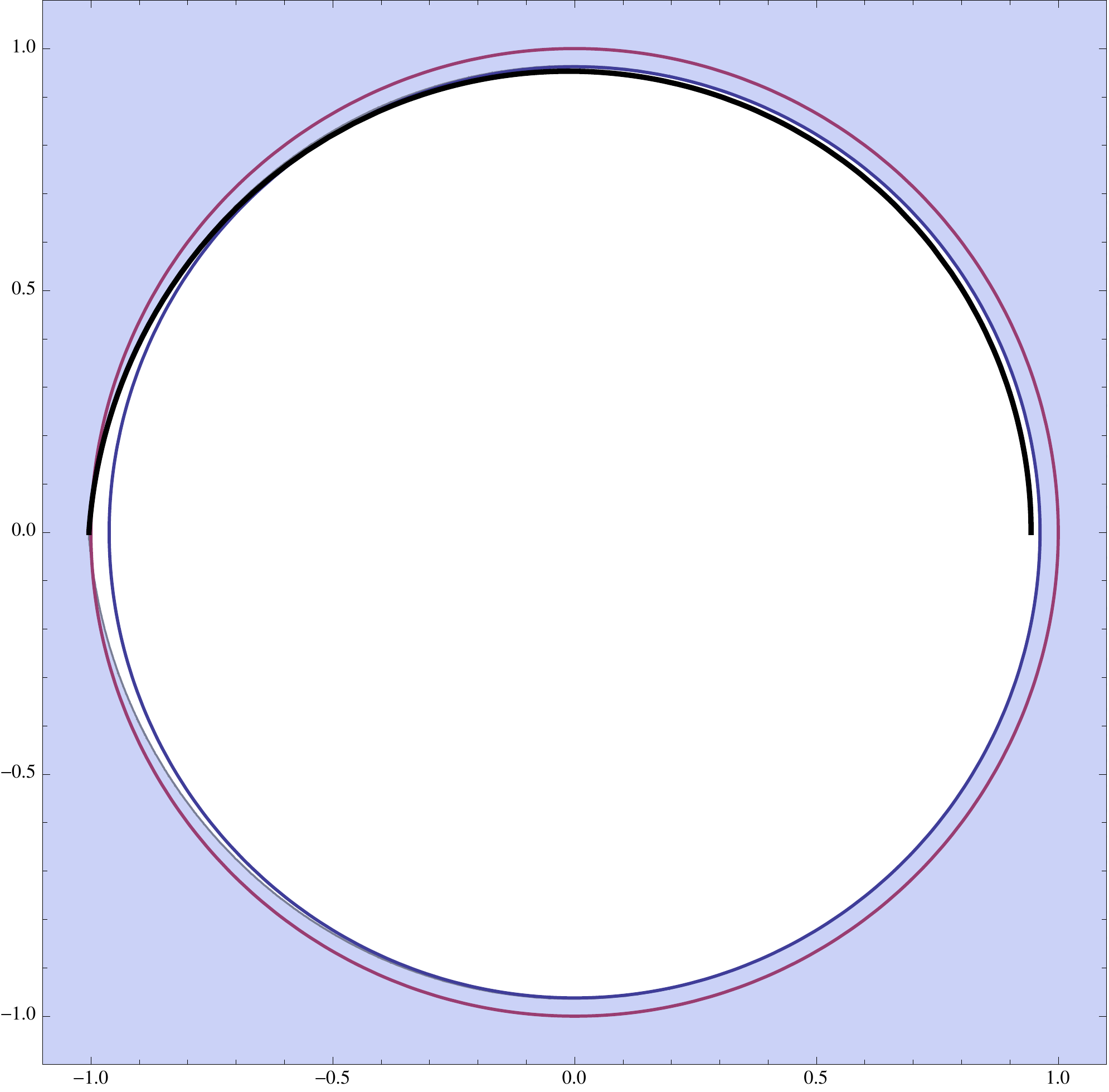}
\end{subfigure}
\end{tabular}
\caption{(color online). $r(\theta)$ (thick black curve), determined by \eqref{eq:thetaR}, plotted against the droplet for $\left|\Delta T\right|=3$ (left) and $5$ (right), showing improvement as $|\Delta T|$ is increased. The concentric circles are $r_s$ (outer ring, red online) and $r_{\ep\rightarrow\infty}$ (inner ring, blue online). The tick marks in the left image are due to a rendering glitch.\label{fig:approxFit}}
\end{figure}

These results can be plotted against the numerical exclusion region (i.e. the droplet) by solving \eqref{eq:Tapprox} for $\ep$, and substituting the result into \eqref{eq:thetaApprox} to obtain an expression for $\Delta\theta(r)$. We find
\begin{align}
\Delta\theta\approx\pi -2 \sqrt{\frac{\left(-1+r+\sqrt{r} \sinh\left(\Delta T/2-\sqrt{r}\right)+\cosh\left(\Delta T/2-\sqrt{r}\right)\right)^2}{2-r-2 \sqrt{r} \sinh\left(\Delta T/2-\sqrt{r}\right)-2 \cosh\left(\Delta T/2-\sqrt{r}\right)}}\label{eq:thetaR}
\end{align}
Two example cases which serve to demonstrate the validity of this result are shown in Fig. \ref{fig:approxFit}. 

We may obtain a more aesthetically pleasing approximation to \eqref{eq:Tapprox} by expanding in the near-horizon region. We find (see appendix)
\begin{empheq}[box=\widefbox]{align}
\pi-\Delta\theta&\approx \sqrt{h}+\frac{1-r}{\sqrt{h}}\label{eq:thetaSimple}
\end{empheq}
where
\begin{align}
\sqrt{h}\equiv2e^{\Delta T/2-1}\label{eq:Dparam}
\end{align}
our choice of the notation ``$\sqrt{h}$'' will become clear shortly. The accuracy of \eqref{eq:thetaSimple} is comparable to \eqref{eq:thetaR} near the horizon (and hence also on the tip for sufficiently large $|\Delta T|$), but cannot be used along the rest of the droplet body. 

At the horizon itself ($r=1$), the second term in \eqref{eq:thetaSimple} vanishes and we obtain an approximation for the angular width of the droplet tip at the Schwarzschild radius as a function of GP time:
\begin{align}
\pi-\Delta\theta\approx \sqrt{h}=2e^{\Delta T/2-1}\label{eq:width}
\end{align}

\begin{figure}[!h]
\centering
\includegraphics[width=0.49\textwidth]{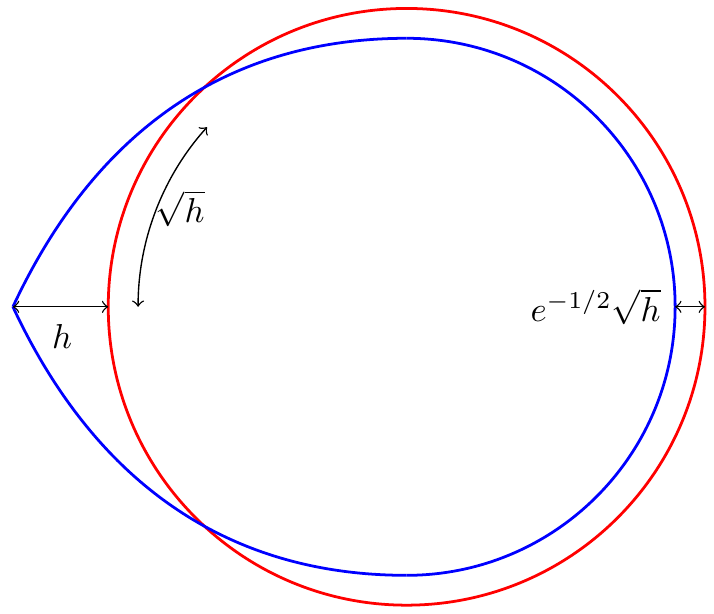}
\caption{(color online). Sketch of a heavily distorted droplet (blue online) against the horizon $r_s=1$ (circle, red online) with parameters of interest labeled: height $h=4e^{\Delta T-2}$, width $\sqrt{h}$, and depth $e^{-1/2}\sqrt{h}$. Note that distances are not to scale, although the height is indeed less than the width for $h<<1$ ($|\Delta T|$ large).\label{fig:sketch}}
\end{figure}

Two other droplet parameters are of interest: the height of the tip above the horizon, and the depth of the antipodal point within. The former is defined by $\Delta\theta=\pi$; hence $\epsilon=\sqrt{r-1}$ and \eqref{eq:Tapprox} becomes
\begin{align}
\Delta T\approx2 \sqrt{r}+2 \log \left(\frac{1-\sqrt{r}}{\sqrt{r-1}}\right)
\end{align}
where we have discarded the negligible third term. Defining the height of the tip $h\equiv r-1>0$, and expanding around $h=0$, we find
\begin{align}
\Delta T\approx2+\frac{h}{2}-\nlog(4)+\nlog(h)
\end{align}
We may then drop the term linear in $h$ relative to the log, and solve:
\begin{align}
h\approx4e^{\Delta T-2}\label{eq:height}
\end{align}
cf. \eqref{eq:Dparam}. To obtain a similar expression for the depth of the antipodal point requires a formula valid in the limit $\epsilon\rightarrow\infty$. From \eqref{eq:largeEpintegrandAppendix} it follows that
\begin{align}
\lim_{\epsilon\rightarrow\infty}\Delta T=2\sqrt{r}+r+2\nlog\left(1-\sqrt{r}\right)
\end{align}
Defining the depth $d\equiv1-r>0$ and expanding, we find
\begin{align}
\Delta T\approx3-\frac{3}{2}d-\nlog(4)+2\nlog(d)
\end{align}
As before, we drop the linear $d$ term and solve:
\begin{align}
d\approx2e^{(\Delta T-3)/2}=e^{-1/2}\sqrt{h}\label{eq:depth}
\end{align}
We summarize our results for the droplet parameters in Fig. \ref{fig:sketch}.

\bigskip

\section{Entropy and information}\label{sec:entropy}
Having obtained a geometric picture of the infalling observer's causal patch in the case of Schwarzschild black holes in $3+1$ dimensions, we now wish to ask what this implies for the AMPS paradox. We appear to have a trade-off between the energy scale of the measurable modes and the angular resolution; i.e., one has large angular visibility only for interior modes that are highly radially localized near the horizon (see Fig. \ref{fig:exclusionregions}). What can we then conclude about the entanglement of the partner modes?

For an infalling observer, the entanglement entropy across the horizon may be thought of as being organized into localized Bell pairs, each of which contains a single bit of entanglement entropy.\cite{Susskind_HH} Consider the total number of bits within the droplet $m=\theta_{\mathrm{missing}}^2/\lambda^2$, out of a total $N=1/\lambda^2$ bits distributed over the entire circle. The wavelength of measurable quanta is limited by the distance between the droplet and the horizon, which for partner modes must be equal inside and outside the black hole. Hence we have $\lambda\le h$ with $\Delta\theta_{\mathrm{missing}}\sim\sqrt{h}$, and therefore
\begin{align}
m=\frac{\left(\Delta\theta_{\mathrm{missing}}\right)^2}{\lambda}\left(\frac{1}{\lambda}\right)\sim\frac{h}{\lambda}\sqrt{N}\implies m\gtrsim\sqrt{N}
\end{align}
where $h/\lambda\ge1$. Thus we find that a single observer is always missing at least about $\sqrt{N}$ out of $N$ bits. Insofar as $N$ is proportional to $\lambda^{-2}$, only high-energy modes stand a chance of reducing the missing fraction to the point where collection of sufficient information is possible. Another obvious though important consequence is that, since one cannot speak of trans-Planckian modes in the absence of a full theory of quantum gravity, $m$ will \emph{never} be zero: even the most determined observer is missing at least one bit.

We may also compute the entropy associated with this missing area. Computing the solid angle in the small $h$ approximation, we find
\begin{align}
A_{\mathrm{missing}}&\approx\pi r^2h\implies\\
S_{\mathrm{missing}}&=\frac{A_{\mathrm{missing}}}{4l_P^2}\approx\frac{\pi r^2}{4l_P^2}h
\end{align}
where we have taken $k_B=1$. Via \eqref{eq:height}, this can be written
\begin{align}
S_{\mathrm{missing}}&\approx\frac{\pi r^2}{l_P^2}e^{\Delta T-2}\label{eq:missingS}
\end{align}
Thus, an observer who wishes to measure a mode with wavelength of order $\lambda\sim h\approx4e^{\Delta T-2}$ does so at an entropy cost given by \eqref{eq:missingS}, which we may think of as the entropy associated with the missing $\sqrt{N}$ bits.

It is interesting to note the consequences for Bousso's double-purity argument\cite{Bousso_purity} in the context of the casual patch considerations above. Essentially, the standard argument is as follows: let $X$ be the interior Hawking quanta, $Y$ the outgoing partner mode, and $Z$ the early Hawking radiation. Then the strong subadditivity of entanglement entropy
\begin{align}
S_{XYZ}+S_Y\le S_{XY}+S_{YZ}\label{eq:SSA}
\end{align}
prevents both $XY$ (the infalling vacuum) and $YZ$ (the final out-state) from being pure. That is, $\forall Z:X\cap Z=\varnothing$,
\begin{align}
S_{XY}=0\land S_Y>0&\implies S_{YZ}\ne0\\
S_{YZ}=0&\implies \nexists X:\left(S_{XY}=0\land S_Y>0\right)
\end{align}
Alternatively, as shown in \cite{Araki_Lieb_1970}, \eqref{eq:SSA} is equivalent to
\begin{align}
S_X+S_Z\le S_{XY}+S_{YZ}\label{eq:SSALaurens}
\end{align}
from which Bousso's conclusion follows immediately. 

However, one can only claim the validity of \eqref{eq:SSA} if one has access to the global field theory. In contrast, here one only has access to some subset of the degrees of freedom, and one can obtain the corresponding entropy inequality within a single causal patch as follows. Define $\tilde{X}\subset X$ as the portion that the infalling observer can see, i.e. $X\equiv\tilde{X}\cup D$ where $D$ is the portion obscured by the droplet at the horizon. Similarly for $\tilde{Y}$. We must also consider that only a small portion of the near-horizon radiation $Y$ ($\tilde{Y}$, $D$) will evolve though the angular momentum barrier to contribute to the late Hawking radiation. Call this subset $Y_R$ ($\tilde{Y}_R$, $D_R$). With these definitions in hand, strong subadditivity can only be formulated in the infalling patch (for the external observer cannot see any of $X$) as
\begin{align}
S_{\tilde{X}\tilde{Y}Z}+S_{\tilde{Y}}\le S_{\tilde{X}\tilde{Y}}+S_{\tilde{Y}Z}\label{eq:SSAin}
\end{align}
and the desired double-purity is really
\begin{align}
S_{\tilde{X}\tilde{Y}}=0\land S_{Y_RZ}=0
\end{align}
with $S_{\tilde{Y}}>0$ and $S_{Y_R}>0$. In contrast to the original argument above, it is by no means obvious that the both the infalling vacuum $\tilde{X}\tilde{Y}$ and the final out-state $Y_RZ$ cannot be pure. That is, when the limitations of the causal patch geometry are taken into account, it may still be possible for both the infalling and external observers to see a pure state without violating the monogamy of entanglement.

An outstanding question is precisely how much of the horizon area---equivalently, how many bits $m$---the infalling observer can afford to lose before measurement of the ingoing Hawking mode becomes impossible. Questions of reconstructing information from some subset of bits are considered in quantum information theory in the context of $(k,n)$ threshold schemes \cite{Cleve_Gottesman_Lo_1999}, in which a quantum ``secret'' is divided into $n$ shares such that any $k\le n$ of those shares can be used to reconstruct the original secret, but any $k-1$ or fewer cannot. The authors of \cite{Cleve_Gottesman_Lo_1999} demonstrated that the only general constraint on such threshold schemes is due to monogamy: one must have $n<2k$ or else the quantum no-cloning theorem is violated. 

Consider, as above, an s-wave immediately behind the horizon with an outgoing partner mode directly outside, with the entanglement information distributed in $N$ localized Bell pairs. Further suppose that the information necessary to reconstruct the entangled state is encoded in a $(k,n)$ threshold scheme ($n=N$). The question at hand is then: what is the value of $k$ needed to reconstruct the state? 

If reconstruction requires the full $N$ bits ($k=n=N$), then our results imply that doing so is impossible, since one reaches the Planck scale in wavelength before the missing number of bits $m\rightarrow0$. Conversely, if the information can be retrieved from some sufficiently large fraction $\frac{N-m}{N}$, then the infalling observer may still be able to reconstruct high-energy modes. In the absence of a precise statement about how black holes encode their secrets, the general bounds $k\le n<2k$ are not sufficiently strict to rule out the possibility that an infalling observer could reconstruct the state despite missing a large number of bits. 

However, this still involves a trade-off between the energy scale of the measurable modes and the angular occlusion. It may be that one can only effect sufficient angular resolution for modes whose energy exceeds some critical value, $\lambda_{\mathrm{crit}}^{-1}$, in which case the $\mathcal{O}(1)$ corrections to high-energy modes purported by AMPS---in contradiction to BHC---would only be detectable for very high-energy modes indeed. More work is needed to determine precisely how small the fraction $m/N$ need be.

\section{Conclusions}
We have shown that for static black holes in $3+1$ and higher dimensions, there does not exist a causal patch that contains all the ingredients necessary to construct the firewall paradox at the level of s-wave Hawking quanta. A possible exception to this principle arises when considering the Schwarzschild black hole in $3+1$ dimensions, and we presented a detailed analysis of the infalling geometry for this case. Our results indicate that the infalling observer is always missing some finite amount of information about the s-wave. Though it remains to show precisely how much angular resolution the observer can afford to lose before reconstruction of the partner mode becomes impossible in principle, our analysis suggests that it is at best difficult in practice.

We focused on the situation for s-waves, as this version of the firewall paradox is the simplest and most robust in our view. Although it would be interesting to consider the consequences for high-$\ell$ modes, this requires a more thorough understanding of the degree to which the mining apparatus disrupts the entanglement of the quantum state. A more detailed analysis of localization of partner modes may shed more light on this direction and we leave that for future work. 

We conclude that for static black holes in $3+1$ and higher dimensions, black hole complementarity is sufficient to evade at least the simplest version of the firewall paradox. Schwarzschild black holes in $3+1$ dimensions nearly allow the paradox to arise within one causal patch, and it is possible that the firewall arguments in that case can be improved, violating complementarity. For rotating black holes and discussions of high-$\ell$ modes using mining, more work is needed.

\section*{Acknowledgements}
We thank Jan de Boer, Raphael Bousso, Irfin Ilgin, Steve Shenker, and Erik Verlinde for discussions. We particularly thank Ted Jacobsen for suggesting the use of the Gullstrand-Painlev\'{e} coordinates, and Douglas Stanford and Lenny Susskind for discussions of the causal patch geometry. This work is part of the $\Delta$-ITP consortium and supported in part by the Foundation for Fundamental Research on Matter (FOM); both are parts of the Netherlands Organization for Scientific Research (NWO) funded by the Dutch Ministry of Education, Culture, and Science (OCW).

\pagebreak
\begin{appendices}
\section{Approximations}
In this appendix we derive the approximate expressions for $\Delta\theta$ \eqref{eq:thetaApprox}, $\Delta T$ \eqref{eq:Tapprox}, and $\Delta\theta\left(r,\Delta T\right)$ \eqref{eq:thetaSimple}. We begin with Eqs. \eqref{eq:theta} and \eqref{eq:T} for the Schwarzschild metric:
\begin{align*}
\Delta\theta&=\int_0^{r_f}\frac{\pm\dr}{\sqrt{\ep^2r^4+r(r_s-r)}}\\
\Delta T&=\int_0^{r_f}\frac{\ep r^3+\sqrt{r_sr}\sqrt{\ep^2 r^4+r(r_s-r)}}{(r_s-r)\sqrt{\ep^2 r^4+r(r_s-r)}}\dr
\end{align*}
where $\ep\in(-\infty,0]$ for $r\ge r_0$, and $\ep\in[0,\infty)$ for $r\le r_0$, with $r_0$ denoting the angular limit where $\ell\rightarrow\infty\implies\ep\rightarrow0$. Note that $r_0<r_s<r_f$, but both $r_0$ and $r_f$ approach $r_s$ asymptotically as $\left|\Delta T\right|$ increases. Note that in our convention, $\Delta T<0$.

Beginning with the $\theta$ integral: for simplicity of notation, consider only the positive case (the negative is merely a mirror image about the $x$-axis). Observe that
\begin{align}
\lim_{\ep\rightarrow\infty}\Delta\theta=0\label{eq:eplimit}
\end{align}
and hence a suitable approximation can be obtained by evaluating the integral for small $\epsilon$.\footnote{This is to be expected, since $\epsilon\rightarrow\infty$ corresponds to the radial limit, in which the angular distance vanishes.} Now suppose there exists an $r'$ such that
\begin{align}
\ep^2r'^4<<r'(r_s-r')\\
r_s-r'<<r_s
\end{align}
Intuitively, the first of these says that the distance to the horizon dominates over the contribution from $\ep$, while the second says that we are still sufficiently close to the horizon that $\ep$ has not yet become large. (These conditions are easily seen to be consistent with the small $\epsilon$ regime, as they can be combined to yield $\epsilon^2r'^3<<r_s$, which for the near-horizon region reduces to $\epsilon<<r_s^{-1}$.)

Thus we can break the integral into two regions:
\begin{align}
\Delta\theta\approx\int_0^{r'}\frac{\dr}{\sqrt{r(r_s-r)}}+\int_{r'}^{r_f}\frac{\dr}{\sqrt{\ep^2r_s^4+r_s(r_s-r)}}\label{eq:thetaSplit}
\end{align}
where in the second term we have expanded to first order in $\delta=r_s-r<<1$. Evaluating \eqref{eq:thetaSplit} yields
\begin{align}
\Delta\theta\approx2\arcsin\left(\sqrt{\frac{r'}{r_s}}\right)-2\sqrt{1-\frac{r_f}{r_s}+\ep^2r_s^2}+2\sqrt{1-\frac{r'}{r_s}+\ep^2r_s^2}\label{eq:thetaEval}
\end{align}
It now remains to eliminate the $r'$ parameter. In the limit that $r'\rightarrow r_s$, $\arcsin\left(\sqrt{\frac{r'}{r_s}}\right)\approx\frac{\pi}{2}-\sqrt{1-\frac{r'}{r_s}}$, and this second term cancels with the last term in \eqref{eq:thetaEval} after dropping the negligible $\ep^2$ contribution. Hence, setting $r_s$ and dropping the subscript on $r_f$,
\begin{align}
\Delta\theta\approx\pi-2\sqrt{1-r+\ep^2}\label{eq:thetaApproxAppendix}
\end{align}
which is \eqref{eq:thetaApprox}.

Turning now to the $T$ integral, we first observe that
\begin{align}
 \lim_{\ep\rightarrow\infty}\Delta T=\int\frac{\sqrt{r}}{\sqrt{r}-1}\dr\label{eq:largeEpintegrandAppendix}
\end{align}
and thus one would not expect the same small $\epsilon$ approximation to suffice for the entire droplet. However, it so happens that the region of large $\epsilon$ is confined relatively close to---that is, has a small angular deviation from---the base of the droplet where $\epsilon\rightarrow\infty$, and as we shall see, the small $\epsilon$ approximation is perfectly adequate elsewhere.

Performing a similar split as in \eqref{eq:thetaSplit} yields
\begin{align}
\Delta T\approx\int_0^{r'}\frac{\sqrt{r_sr}}{r-r_s}\dr+\int_{r'}^{r_f}\frac{\ep r_s^3+r_s\sqrt{\ep^2r_s^4+r_s(r_s-r)}}{(r-r_s)\sqrt{\ep^2r_s^4+r_s(r_s-r)}}\dr
\end{align}
where the second term has again been expanded to first order in the near-horizon region. Rather than integrate immediately however, we first analytically eliminate the $r'$ parameter by extending the integration regions and subtracting compensating terms:
\begin{align*}
\Delta T\approx&\int_0^{r_f}\frac{\sqrt{r_sr}}{r-r_s}\dr+\int_0^{r_f}\frac{\ep r_s^3+r_s\sqrt{\ep^2r_s^4+r_s(r_s-r)}}{(r-r_s)\sqrt{\ep^2r_s^4+r_s(r_s-r)}}\dr\\
-&\int_{r'}^{r_f}\frac{\sqrt{r_sr}}{r-r_s}\dr-\int_0^{r'}\frac{\ep r_s^3+r_s\sqrt{\ep^2r_s^4+r_s(r_s-r)}}{(r-r_s)\sqrt{\ep^2r_s^4+r_s(r_s-r)}}\dr
\end{align*}
Note that the third term is now entirely in the region where $r_s\sim r$, while the fourth is in the regime where $r_s-r$ dominates over $\ep$. Hence:
\begin{align*}
\Delta T\approx&\int_0^{r_f}\frac{\sqrt{r_sr}}{r-r_s}\dr+\int_0^{r_f}\frac{\ep r_s^3+r_s\sqrt{\ep^2r_s^4+r_s(r_s-r)}}{(r-r_s)\sqrt{\ep^2r_s^4+r_s(r_s-r)}}\dr-\int_{r'}^{r_f}\frac{r_s}{r-r_s}\dr-\int_0^{r'}\frac{r_s}{r-r_s}\dr\\
=&\int_0^{r_f}\frac{\sqrt{r_sr}}{r-r_s}\dr+\int_0^{r_f}\frac{\ep r_s^3+r_s\sqrt{\ep^2r_s^4+r_s(r_s-r)}}{(r-r_s)\sqrt{\ep^2r_s^4+r_s(r_s-r)}}\dr-\int_0^{r_f}\frac{r_s}{r-r_s}\dr
\end{align*}
The advantage of this seemingly roundabout exercise is that now each of the above terms can be integrated indefinitely, and the result rearranged prior to plugging in limits in order to obtain a finite, real result. Setting $r_s$ to 1 for simplicity, we find
\begin{align*}
\Delta T&\approx2\sqrt{r}-2\arctanh(\sqrt{r})-2\arctanh\left(\frac{1}{\ep}\sqrt{1-r+\ep^2}\right)+\nlog(1-r)-\nlog(r-1)\\
&=2\sqrt{r}-\nlog\left(\frac{1+\sqrt{r}}{1-\sqrt{r}}\right)-\nlog\left(\frac{\ep+\sqrt{1-r+\ep^2}}{\ep-\sqrt{1-r+\ep^2}}\right)+\nlog(1-r)-\nlog(r-1)\\
&=2\sqrt{r}-\nlog\left(\frac{1+\sqrt{r}}{1-\sqrt{r}}\right)-2\nlog\left(\ep+\sqrt{1-r+\ep^2}\right)\\
&+\nlog\left((\ep-\sqrt{1-r+\ep^2})(\ep+\sqrt{1-r+\ep^2})\right)+\nlog(1-r)-\nlog(r-1)\\
&=2\sqrt{r}-\nlog\left(\frac{1+\sqrt{r}}{1-\sqrt{r}}\right)-2\nlog(\ep+\sqrt{1-r+\ep^2})+\nlog\left((1+\sqrt{r})(1-\sqrt{r})\right)\\
&=2\sqrt{r}+2\nlog(1-\sqrt{r})-2\nlog(\ep+\sqrt{1-r+\ep^2})\\
&=2\sqrt{r}+2\nlog\left(\frac{1-\sqrt{r}}{\ep+\sqrt{1-r+\ep^2}}\right)\\
\end{align*}
Thus, plugging in limits of integration (again dropping the subscript on $r_f$),
\begin{align}
\Delta T=2\sqrt{r}+2\nlog\left(\frac{1-\sqrt{r}}{\ep+\sqrt{1-r+\ep^2}}\right)+2\nlog\left(\ep+\sqrt{1-\ep^2}\right)\label{eq:TapproxAppendix}
\end{align}
which is \eqref{eq:Tapprox}. 

Eliminating $\epsilon$ in order to combine \eqref{eq:thetaApproxAppendix} and \eqref{eq:TapproxAppendix} leads to the full expression for the droplet body given in the main text, \eqref{eq:thetaR}. Here we obtain a simpler expression, which is still reasonably accurate away from the droplet base where $\epsilon$ becomes large. Defining $x\equiv1-r$ in the small-$\epsilon$ regime, we have, to first order,
\begin{align}
&\Delta T\approx2-x+2\nlog\left(\frac{x}{2(\epsilon+\sqrt{x+\epsilon^2})}\right)+2\epsilon\implies\notag\\
&\frac{\Delta T}{2}-1\approx\nlog\left(\frac{x}{2(\epsilon+\sqrt{x+\epsilon^2})}\right)\label{eq:Tintermediate}
\end{align}
where we have dropped the linear $x<<1$ and $\epsilon<<1$ terms. Then, from \eqref{eq:thetaApprox}, we have
\begin{align}
2\sqrt{x+\epsilon^2}\approx\pi-\Delta\theta\equiv\alpha\implies\epsilon\approx\sqrt{\alpha^2/4-x}
\end{align}
which we substitute into \eqref{eq:Tintermediate}:
\begin{align*}
xe^{-(\Delta T/2-1)}\approx\sqrt{\alpha^2-4x}+\alpha&\implies
2\alpha e^{-(\Delta T/2-1)}\approx xe^{-2(\Delta T/2-1)}+4\implies\notag\\
\alpha&\approx\frac{x}{2}e^{-(\Delta T/2-1)}+2e^{\Delta T/2-1}
\end{align*}
Thus we obtain
\begin{align}
\pi-\Delta\theta&\approx \sqrt{h}+\frac{x}{\sqrt{h}}\label{eq:thetaSimpleAppendix}
\end{align}
where
\begin{align}
x\equiv1-r\mathrm{,}\;\;\;&\;\;\;\sqrt{h}\equiv2e^{\Delta T/2-1}\label{eq:DparamAppendix}
\end{align}
which is \eqref{eq:thetaSimple}. As shown in the main text, the notation ``$\sqrt{h}$'' was chosen so as to write the various droplet parameters in terms of the height $h$ of the tip above the horizon.
\end{appendices}

\bibliographystyle{arXiv}
\bibliography{geomfire}

\begin{thebibliography}{10}
\newcommand{\enquote}[1]{``#1''}
\providecommand{\url}[1]{\texttt{#1}}
\providecommand{\urlprefix}{URL }
\expandafter\ifx\csname urlstyle\endcsname\relax
  \providecommand{\doi}[1]{doi:\discretionary{}{}{}#1}\else
  \providecommand{\doi}{doi:\discretionary{}{}{}\begingroup
  \urlstyle{rm}\Url}\fi
\providecommand{\bibinfo}[2]{#2}
\providecommand{\eprint}[2][]{\url{#2}}

\bibitem{AMPS}
\bibinfo{author}{A.~Almheiri et~al.}, \enquote{\bibinfo{title}{Black holes:
  complementarity or firewalls?}}   (\bibinfo{year}{2013}),
  \bibinfo{note}{\href{http://arxiv.org/abs/1207.3123}{arXiv:1207.3123
  [hep-th]}}.

\bibitem{Hawking_1976}
\bibinfo{author}{S.~W. Hawking}, \enquote{\bibinfo{title}{Breakdown of
  predictability in gravitational collapse},} \bibinfo{journal}{Phys. Rev. D}
  \textbf{\bibinfo{volume}{14}}, \bibinfo{number}{2460} (\bibinfo{year}{1976}),
  \bibinfo{note}{\href{http://journals.aps.org/prd/abstract/10.1103/PhysRevD.1%
4.2460}{PhysRevD.14.2460}}.

\bibitem{Mat09}
\bibinfo{author}{S.~D. Mathur}, \enquote{\bibinfo{title}{The information
  paradox: A pedagogical introduction},} \bibinfo{journal}{Class. Quant. Grav.}
  \textbf{\bibinfo{volume}{26}}:\bibinfo{pages}{224001} (\bibinfo{year}{2009}),
  \doi{10.1088/0264-9381/26/22/224001},
  \bibinfo{note}{\href{http://arXiv.org/abs/0909.1038}{arXiv:0909.1038
  [hep-th]}}.

\bibitem{BraPir09}
\bibinfo{author}{S.~L. Braunstein}, \bibinfo{author}{S.~Pirandola}, and
  \bibinfo{author}{K.~\'{Z}yczkowski}, \enquote{\bibinfo{title}{Entangled black
  holes as ciphers of hidden information},} \bibinfo{journal}{Phys. Rev. Lett.}
  \textbf{\bibinfo{volume}{110}}:\bibinfo{pages}{101301}
  (\bibinfo{year}{2013}),
  \bibinfo{note}{\href{http://arXiv.org/abs/0907.0739}{arXiv:0907.1190
  [quant-ph]}}.

\bibitem{BHC}
\bibinfo{author}{L.~Susskind}, \bibinfo{author}{L.~Thorlacius}, and
  \bibinfo{author}{J.~Uglum}, \enquote{\bibinfo{title}{The stretched horizon
  and black hole complementarity},} \bibinfo{journal}{Phys. Rev. D}
  \textbf{\bibinfo{volume}{48}}, \bibinfo{number}{3743} (\bibinfo{year}{1993}),
  \bibinfo{note}{\href{http://arxiv.org/abs/hep-th/9306069}{arXiv:hep-th/93060%
69}}.

\bibitem{Ilgin_Yang_2013}
\bibinfo{author}{I.~Ilgin} and \bibinfo{author}{I-S. Yang},
  \enquote{\bibinfo{title}{Causal patch complementarity: The inside story for
  old black holes},}   (\bibinfo{year}{2013}),
  \bibinfo{note}{\href{http://arXiv.org/abs/1311.1219}{arXiv:1311.1219
  [hep-th]}}.

\bibitem{Freivogel_2014}
\bibinfo{author}{B.~Freivogel}, \enquote{\bibinfo{title}{Energy and information
  near black hole horizons},}   (\bibinfo{year}{2014}),
  \bibinfo{note}{\href{http://arXiv.org/abs/1401.5340}{arXiv:1401.5340v1
  [hep-th]}}.

\bibitem{Marolf_Polchinski_2013}
\bibinfo{author}{D.~Marolf} and \bibinfo{author}{J.~Polchinski},
  \enquote{\bibinfo{title}{Gauge/gravity duality and the black hole interior},}
  \bibinfo{journal}{Phys. Rev. Lett.} \textbf{\bibinfo{volume}{111}},
  \bibinfo{number}{171301} (\bibinfo{year}{2013}),
  \bibinfo{note}{\href{http://arxiv.org/abs/1307.4706}{arXiv:1307.4706
  [hep-th]}}.

\bibitem{AMPSS}
\bibinfo{author}{A.~Almheiri et~al.}, \enquote{\bibinfo{title}{An apologia for
  firewalls},}   (\bibinfo{year}{2013}),
  \bibinfo{note}{\href{http://arXiv.org/abs/1304.6483}{arXiv:1304.6483
  [hep-th]}}.

\bibitem{HH}
\bibinfo{author}{D.~Harlow} and \bibinfo{author}{P.~Hayden},
  \enquote{\bibinfo{title}{Quantum computation vs. firewalls},}
  \bibinfo{journal}{JHEP} \textbf{\bibinfo{volume}{1306}}:\bibinfo{pages}{085}
  (\bibinfo{year}{2013}), \doi{10.1007/JHEP06(2013)085},
  \bibinfo{note}{\href{http://arXiv.org/abs/1301.4504}{arXiv:1301.4504
  [hep-th]}}.

\bibitem{HuiYan13}
\bibinfo{author}{L.~Hui} and \bibinfo{author}{I-S. Yang},
  \enquote{\bibinfo{title}{{Complementarity + Back-reaction is enough}},}
  (\bibinfo{year}{2013}),
  \bibinfo{note}{\href{http://arXiv.org/abs/1308.6268}{arXiv:1308.6268
  [hep-th]}}.

\bibitem{Bousso_etal_2013}
\bibinfo{author}{R.~Bousso et~al.}, \enquote{\bibinfo{title}{Null geodesics,
  local {CFT} operators and {AdS/CFT} for subregions},} \bibinfo{journal}{Phys.
  Rev. D} \textbf{\bibinfo{volume}{88}}, \bibinfo{number}{064057}
  (\bibinfo{year}{2013}),
  \bibinfo{note}{\href{http://arXiv.org/abs/1209.4641}{arXiv:1209.4641
  [hep-th]}}.

\bibitem{Leichenauer_Rosenhaus_2013}
\bibinfo{author}{S.~Leichenauer} and \bibinfo{author}{V.~Rosenhaus},
  \enquote{\bibinfo{title}{{AdS} black holes, the bulk-boundary dictionary, and
  smearing functions},} \bibinfo{journal}{Phys. Rev. D}
  \textbf{\bibinfo{volume}{88}}, \bibinfo{number}{2} (\bibinfo{year}{2013}),
  \bibinfo{note}{\href{http://arXiv.org/abs/1304.6821}{arXiv:1304.6821
  [hep-th]}}.

\bibitem{Rosenhaus_Rey_2014}
\bibinfo{author}{S-J. Rey} and \bibinfo{author}{V.~Rosenhaus},
  \enquote{\bibinfo{title}{Scanning tunneling macroscopy, black holes, and
  {AdS/CFT} bulk locality},}   (\bibinfo{year}{2014}),
  \bibinfo{note}{\href{http://arXiv.org/abs/1403.3943}{arXiv:1403.3943
  [hep-th]}}.

\bibitem{Brown_2012}
\bibinfo{author}{A.~Brown}, \enquote{\bibinfo{title}{Tensile strength and the
  mining of black holes},}   (\bibinfo{year}{2012}),
  \bibinfo{note}{\href{http://arXiv.org/abs/1207.3342}{arXiv:1207.3342v1
  [gr-qc]}}.

\bibitem{Bousso_Freivogel_Leichenauer_2010}
\bibinfo{author}{R.~Bousso}, \bibinfo{author}{B.~Freivogel}, and
  \bibinfo{author}{S.~Leichenauer}, \enquote{\bibinfo{title}{Saturating the
  holographic entropy bound},} \bibinfo{journal}{Phys. Rev. D}
  \textbf{\bibinfo{volume}{82}}, \bibinfo{number}{084024}
  (\bibinfo{year}{2010}),
  \bibinfo{note}{\href{http://arxiv.org/abs/1003.3012}{arXiv:1003.3012
  [hep-th]}}.

\bibitem{Kabat:2011rz}
\bibinfo{author}{D.~Kabat}, \bibinfo{author}{G.~Lifschytz}, and
  \bibinfo{author}{D.~A. Lowe}, \enquote{\bibinfo{title}{Constructing local
  bulk observables in interacting {AdS/CFT}},} \bibinfo{journal}{Phys. Rev. D}
  \textbf{\bibinfo{volume}{83}}:\bibinfo{pages}{106009} (\bibinfo{year}{2011}),
  \doi{10.1103/PhysRevD.83.106009},
  \bibinfo{note}{\href{http://arXiv.org/abs/1102.2910}{arXiv:1102.2910
  [hep-th]}}.

\bibitem{Martel_Poisson_2000}
\bibinfo{author}{K.~Martel} and \bibinfo{author}{E.~Poisson},
  \enquote{\bibinfo{title}{Regular coordinate systems for schwarzschild and
  other spherical spacetimes},}   (\bibinfo{year}{2000}),
  \bibinfo{note}{\href{http://arxiv.org/abs/gr-qc/0001069}{arXiv:gr-qc/0001069%
}}.

\bibitem{Susskind_HH}
\bibinfo{author}{L.~Susskind}, \enquote{\bibinfo{title}{Black hole
  complementarity and the hawlow-hayden conjecture},}   (\bibinfo{year}{2013}),
  \bibinfo{note}{\href{http://arXiv.org/abs/1301.4505}{arXiv:1301.4505v2
  [hep-th]}}.

\bibitem{Bousso_purity}
\bibinfo{author}{R.~Bousso}, \enquote{\bibinfo{title}{Firewalls from double
  purity},} \bibinfo{journal}{Phys. Rev. D} \textbf{\bibinfo{volume}{88}},
  \bibinfo{number}{084035} (\bibinfo{year}{2013}),
  \bibinfo{note}{\href{http://arxiv.org/abs/1308.2665}{arXiv:1308.2665
  [hep-th]}}.

\bibitem{Araki_Lieb_1970}
\bibinfo{author}{H.~Araki} and \bibinfo{author}{E.~H. Lieb},
  \enquote{\bibinfo{title}{Entropy inequalities},} \bibinfo{journal}{Commun.
  Math. Phys.} \textbf{\bibinfo{volume}{18}}:\bibinfo{pages}{160--170}
  (\bibinfo{year}{1970}),
  \bibinfo{note}{\href{http://projecteuclid.org/euclid.cmp/1103842506}{Project
  Euclid}}.

\bibitem{Cleve_Gottesman_Lo_1999}
\bibinfo{author}{R.~Cleve}, \bibinfo{author}{D.~Gottesman}, and
  \bibinfo{author}{H-K. Lo}, \enquote{\bibinfo{title}{How to share a quantum
  secret},} \bibinfo{journal}{Phys. Rev. Lett.} \textbf{\bibinfo{volume}{83}},
  \bibinfo{number}{648} (\bibinfo{year}{1999}),
  \bibinfo{note}{\href{http://arxiv.org/abs/quant-ph/9901025}{arXiv:quant-ph/9%
901025}}.

\end{thebibliography}
\end{document}